\documentclass[%
 reprint,
 amsmath,amssymb,
 aps,
]{revtex4-2}

\usepackage{graphicx}
\usepackage{dcolumn}
\usepackage{bm}

\usepackage[usenames,dvipsnames]{color}
\usepackage{hyperref}
\usepackage{amsmath}
\usepackage{comment}
 \usepackage[normalem]{ulem}
 \newcommand{\sqsn}{$\sqrt{s_{_{NN}}}$}

\begin{document}

\preprint{APS/123-QED}

\title{Bjorken Initial Energy Density and Viscous Longitudinal Hydrodynamic Evolution in 
Xe-Xe Collisions}
\author{S. Biswal$^{1}$}
\author{M. A. Bhat$^{2}$}
\email{asifqadir1994@gmail.com}
\author{A. Nayak$^{2,3}$}
\author{S. I. Sahoo$^{1, 4}$}
\author{D. Dutta$^{1,4}$}
\author{D. K. Mishra$^{1,4}$}
\author{P. K. Sahu$^{1,2}$}
\email{pradip@iopb.res.in}
\affiliation{${}^1$Homi Bhabha National Institute, Anushakti Nagar, Mumbai 400094, India.\\ 
${}^2$Institute of Physics, Sachivalaya Marg, Sainik School P.O., Bhubaneswar 751005, India.\\
${}^3$Department of Physics, School of Applied Sciences, KIIT, Bhubaneswar 751024, India.\\
${}^4$Nuclear Physics Division, Bhabha Atomic Research Centre, Mumbai 400085, India.}

\begin{abstract}
We present a systematic study of the Bjorken initial energy density in Xe-Xe collisions at 
$\sqrt{s_{_{NN}}} = 5.44$ TeV, estimated using charged-particle multiplicity data and a generalized transverse overlap geometry applicable beyond the most central collisions. 
The dependence of the extracted energy density is examined by adopting both a constant formation time and a centrality-dependent formation time derived from Pb–Pb collisions at $\sqrt{s_{NN}} = 5.02$ TeV.
Corresponding Bjorken energy density estimates for Pb-Pb collisions are also presented for comparison. Taking the Bjorken energy density and formation time as initial conditions, the 
subsequent longitudinal evolution of the quark-gluon plasma (QGP) formed in these 
collisions is studied. Both ideal and first-order viscous boost-invariant hydrodynamics are 
employed to assess the influence of dissipation. We observe that viscous effects slow the longitudinal expansion and lead to entropy production dominated by early-time dynamics. The lifetime of the QGP is observed to increase with centrality and is substantially enhanced by viscous effects. These effects are highly sensitive to the choice of formation time, particularly in peripheral 
collisions. A comparative analysis of Xe-Xe and Pb-Pb collisions demonstrates that the longitudinal evolution is primarily controlled by the initial energy density scale 
set by the Bjorken prescription. Consequently, when this scale is comparable, both systems exhibit nearly identical evolution patterns, while appreciable distinctions emerge in peripheral collisions due to system-size and geometric effects. 
\end{abstract}

\maketitle


\section{\label{A}Introduction}
In the early universe shortly after the Big Bang, a new state of matter is believed 
to have existed called Quark-Gluon Plasma 
(QGP)~\cite{Shuryak:2014zxa,Rafelski:1982pu,Koch:1986ud, 
Koch:1988nn}. 
The existence of QGP in heavy-ion collisions was first proposed in the 
1970s~\cite{Shuryak:1978ij}, an attempt to understand the behavior of high energy 
collisions between atomic nuclei.

Experimental evidence for the creation of the QGP was provided by the Relativistic Heavy Ion Collider (RHIC) through heavy-ion collisions, such as Au-Au interactions, at a center-of-mass energy of
$\sqrt{s_{_{NN}}}$ = 200 GeV~\cite{PHENIX:2004vcz,Bellwied:2005kq,Muller:2006ee}. These 
collisions produced a perfect fluid which is made up of quarks and gluons having the 
ratio of shear viscosity to entropy density ($\eta/s$) smaller than that of any other known 
liquid~\cite{Gyulassy:2004zy,Pal:2010es, 
Tannenbaum:2006ch,Song:2007fn}. The Large Hadron Collider Experiment 
(LHC) later confirmed its existence in Pb-Pb 
collisions~\cite{ALICE:2010suc,ATLAS:2011ah}. In other collision 
systems such as pp, p-Pb at 
LHC~\cite{Kharzeev:2014pha,ALICE:2016fzo,CMS:2016fnw,Bjorken:2013boa} and $^{3}$He-Au, $d$--Au at RHIC~\cite{PHENIX:2018lia} similar 
properties also have been observed.

The most compelling theoretical evidence comes from lattice QCD, which predicts a
phase transition at a critical temperature of approximately 154 MeV ~\cite{Bazavov:2011nk}.
This corresponds to a critical energy density ($\varepsilon$) of about 
0.7–1.3 GeV/fm$^3$, at zero net-baryon density~\cite{Laermann:2003cv}. Estimates 
derived from charged-particle pseudo-rapidity distributions and elliptic flow 
measurements in central Au-Au collisions at $\sqrt{s_{_{NN}}}= 200$  GeV at RHIC 
indicate that, the energy densities ($\varepsilon$) $\geq$ 3 GeV/fm$^3$ are achieved 
as the system approaches approximate equilibrium~\cite{PHOBOS:2004zne}. These 
measurements are about 15$-$20 times higher than the normal nuclear matter density. 
It is therefore essential to determine the extent of the 
space–time region in which a heavy-ion collision exceeds the critical energy density, 
as well as the degree of equilibration attended within that region. Consequently, 
the evolution of the energy density during the early stages of the collisions, 
within the first few fm, are of particular importance.

The evolution of the QGP fluid can be better understood through transport 
coefficients such as viscosity and entropy, as they are closely related to the 
underlying quark-gluon strong interactions~\cite{Yang:2023apw}. On focusing the 
temperature dependence of viscosities, the theoretical calculations on the shear 
($\eta$) and bulk ($\zeta$) viscosity have been extensively 
explored~\cite{Meyer:2007ic,Grefa:2022sav} at vanishing baryon density or 
equivalently zero baryon chemical potential.
The final observables of 
heavy-ion collisions~\cite{Teaney:2003kp,Ryu:2015vwa} are significantly affected by 
the viscosities of the QGP; therefore, their values can be constrained with the 
help of experimental data. In earlier studies of viscous hydrodynamics,
$\frac{\eta}{s} = 0.08 \sim 0.2$~\cite{Romatschke:2007mq,Niemi:2015qia} was 
typically assumed to be constant throughout the entire evolution. Recently, the 
constraints on the temperature dependence of the shear and bulk viscosities of the 
baryon-free QGP~\cite{Bernhard:2015hxa,Heffernan:2023utr} have been obtained using 
Bayesian statistical analysis with multi-stage models which 
integrate initial conditions, viscous hydrodynamics and hadronic transport. These 
studies consistently find an overall value of $\frac{\eta}{s} \approx 0.16$ for the 
baryon-free QGP near pseudo-critical temperature T$_{pc} \approx$ 160 
MeV~\cite{Nijs:2022rme,Parkkila:2021yha}.

Entropy is generated in a viscous evolution, leading to a difference between the 
initial and final state entropies. In Au-Au collisions, explicit numerical 
simulations indicate that, in fluid evolution with viscosity to entropy ratio 
$\frac{\eta}{s} = 0.08, 0.12$ and $0.16$, the corresponding entropy production can 
be 20\%, 30\%, and 50\%, respectively higher than the initial entropy~\cite{Chaudhuri:2009uk}. These simulations 
further indicate that the entropy production in viscous hydrodynamics occurs 
rapidly, with the majority generated within the first $2-4$ fm of 
evolution~\cite{Chaudhuri:2010in}.

The evolution of the system created in ultra-relativistic heavy-ion collisions is 
driven by pressure 
gradients, expands collectively and cools until hadronization. The geometric 
anisotropies in the initial state is turned by the gradient driven expansion into 
anisotropic flow in the final state and the variations in size in the initial state 
into radial flow~\cite{Prasad:2022zbr,Bozek:2012fw,Samanta:2023amp}. The azimuthal 
momentum anisotropy is quantified by the anisotropic 
flow~\cite{ALICE:2011ab,PHENIX:2011yyh,CMS:2013wjq,ATLAS:2019peb}, while the radial boost of the system, 
which influences the average transverse momentum of particles in each event is 
characterized by radial 
flow~\cite{Heinz:2013th,Broniowski:2009fm}.

It is not straight-forward to directly estimate the initial pressure and 
energy density in the experiment. Therefore, hydrodynamical calculations, which are 
based on the assumptions of local thermal equilibrium provide indirect constraints 
on these quantities. In the hydrodynamic framework, a proper time of $\tau_0 \sim$ 1 
fm after the collisions is used as input for the subsequent evolution of the 
system formed in the heavy-ion collisions. Therefore, hydrodynamical model requires 
the initial energy density and net-baryon density, which are commonly assumed to be 
boost-invariant over a broad range of space-time rapidity.

The reaction dynamics can be better understood by the fluid hydrodynamics. In a 
collision at non-zero impact parameter, the anisotropy of the low $p_{T}$ particles 
produced is described by the elliptic flow. This suggests that a collective flow of 
the particles exist following a hydrodynamical pressure gradient which is due to the 
initial eccentricity in a 
collision~\cite{Ollitrault:1992bk,Yadav:2025vtc,Shen:2011eg}. The elliptic 
flow~\cite{Hirano:2008hy} is successfully described by most of the hydrodynamical 
simulations which are compatible with an almost ``perfect fluid'' behavior, i.e. a 
small ratio of $\eta/s$~\cite{Huovinen:2006jp,Niemi:2011ix,Demir:2008tr}.

The hydrodynamical description of the QGP medium created is validated by 
assuming the quasi-perfect fluid behavior~\cite{Daher:2024vxk,Capellino:2022nvf}. In 
the reaction process hypothesis, an intermediate stage, such as a boost-invariant QGP phase modeled as a relativistically expanding fluid, which 
constitutes the foundation of the Bjorken flow. 
In the central region of the collision, boost-invariance is justified by the 
flatness of the observed particle distribution. This observation agrees with 
hydrodynamic predictions of boost invariance, in which the fluid (space–time) 
rapidity equals with the particle (energy–momentum) rapidity~\cite{Bjorken:1982qr}.

In this work, we perform a systematic study of the Bjorken initial energy density. 
A generalized elliptic transverse overlap geometry is considered that remains valid from central to peripheral collisions. Using the estimated initial conditions, we study the subsequent longitudinal expansion of the QGP medium within the framework of boost-invariant relativistic hydrodynamics. Both ideal and first-order viscous hydrodynamics are investigated to check the effects of viscosity on the energy density evolution, entropy production, and lifetime of the QGP. \par
The paper is organized as follows. The formulation of the Bjorken model is 
discussed in Sec.~\ref{B}. The overlapping area calculation and initial condition 
estimations are done in Sec.~\ref{3A} and~\ref{3B}, respectively. In Sec.~\ref{hy}, we 
present the hydrodynamic evolution of the system. The resulting energy density 
evolution, viscous effects, entropy production, and QGP lifetime are analyzed in 
detail in Sec.~\ref{D}. At the end of Sec.~\ref{D}, a comparative study of Xe-Xe and Pb-Pb collisions is presented to highlight system-size and collision energy 
dependence. Finally, the main conclusions are summarized in Sec.~\ref{E}. 

\section{\label{B}Bjorken Model}
The hydrodynamic description introduced by Bjorken is one of the principal models 
used to describe the heavy-ion collisions at high energies~\cite{Bjorken:1982qr}. 
The popular Bjorken 
flow, which is the 
boost invariant fluid flow does not depend on rapidity but only on proper time 
$(\tau)$~\cite{Gubser:2012gy,Bagchi:2023ysc}. In high-energy heavy-ion collisions at central rapidity, this 
phenomenological 
assumption holds true. This model has been very successful in describing the extreme 
regime, where velocity of fluid is close to the light velocity and is one of the 
simplest models. The space-time evolution of highly energetic and dense state of 
matter created in heavy-ion collisions is described by the Bjorken flow as an 
ultra-relativistic fluid~\cite{Ciambelli:2018xat,Petkou:2022bmz}. All the 
interesting 
dynamics takes place along the direction in which the collision of the two heavy 
nuclei occur i.e. along the beam axis, which is usually taken as $z$-axis. This is 
considered as one among the major simplifying assumptions of Bjorken flow. In the 
transverse $x-y$ plane, the flow expects complete rotational and translational 
invariance, which leads it to become effectively two dimensional. The boost or to be more precise rapidity invariance, which claims about the 
existence of the velocity profile of the produced fluid after the collision, is the 
second major assumption of the Bjorken flow. The longitudinal velocity of the fluid 
at the location $z$ is given by $v = \frac{z}{t}$, after assuming that the collision 
occurred at $z = 0$ and at time $t = 0$. However, at any later time $t$ the fluid 
exactly in the midway between the two receding nuclei continues to be at rest. This 
can also be interpreted as, the fluid at $z = 0$ is at rest at a particular instant 
in time $t$, whereas the fluid is moving with the speed of light at $z = \pm t$. 
Assuming the QGP medium as a fluid system,  the initial 
energy density is given by~\cite{Bjorken:1982qr}:
\begin{equation}
	\varepsilon_{B} =\; \frac{3}{2}\frac{dN_{ch}}{dy}\frac{\left<m_{T}\right>}
	{\tau_{o}A_{\text{overlap}}}\;\;\;\;\;\;\;\;,
    \label{eq:eneden}
\end{equation}
where $\left<m_{T}\right>$ is the transverse mass of the produced particles, 
$\frac{dN_{ch}}{dy}$ being the charged particle rapidity density measured in 
relativistic heavy-ion collisions, $\tau_{0}$ is the formation time of possible 
hydrodynamical system or initial time of a possible hydrodynamic evolution. 
$A_{\text{overlap}}$ is the area of transverse overlapping region. 
According to this model an initial energy density $\varepsilon_{B} \ge$ 1 
GeV/fm$^{3}$ ensures the existence of the QGP medium.\

\section{\label{C} Methodology}
\subsection{Area of Overlapping Region}\label{3A}
\begin{figure}[htb]
    \centering
    \includegraphics[width=0.8\linewidth]{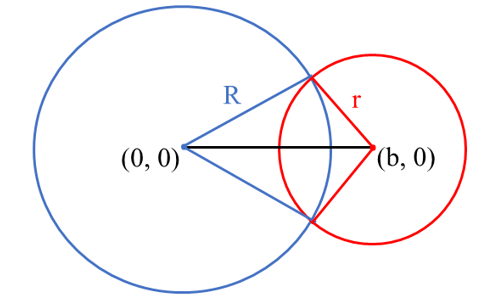}
    \caption{Elliptic transverse overlap geometry formed in a non-central 
    collision between two nuclei of radii $R$ and $r$ at impact parameter $b$.}
    \label{overlaparea}
\end{figure}

In the Bjorken picture, the transverse overlap area is conventionally approximated 
as a circular region, which is applicable for the most central collisions. To study 
with mid-central and most-peripheral collisions, the overlapping region is generalized by 
considering it as an elliptic geometry as shown in Fig.~\ref{overlaparea}.

The generalized overlapping area is calculated as,
\begin{multline}
    A_{\text{overlap}} =\; R^2 \cos^{-1} \left(\frac{b^2-r^2+R^2}{2bR}\right) \\
    + r^2 \cos^{-1} \left(\frac{b^2+r^2-R^2}{2br}\right) \\
    -\frac{1}{2} \sqrt{\left(b+r+R\right)\left(b-r+R\right)
    \left(r+b-R\right)\left(r-b+R\right)}\;\;,
\end{multline}
where $R$, $r$ are the radii of the participating nuclei and $b$ is the 
impact parameter. The radius of the Xe nucleus is calculated as $R = R_0A^{1/3}$, 
where $R_0$ is the Fermi radius ($\sim$ 1.2 fm) and $A$ is the mass number 
of the 
nucleus.
This expression has been derived in detail in Appendix \ref{AA}. As a special case 
of this, if we take collision between two identical nuclei, i.e. $R=r$, then we get
\begin{equation}\label{eq:overlap}
    A_{\text{overlap}} =\; 2R^2 \cos^{-1} \left(\frac{b}{2R}\right)-\frac{b}{2}\sqrt{4R^2-b^2}.
\end{equation}
The consistency of Eq.~(\ref{eq:overlap}) can be checked by the  fact that it 
correctly 
reproduces the limiting cases of complete overlap and vanishing overlap at $b=0$ and 
$b=2R$, respectively.\par
\begin{figure}[htb]
    \centering
    \includegraphics[width=1.0\linewidth]{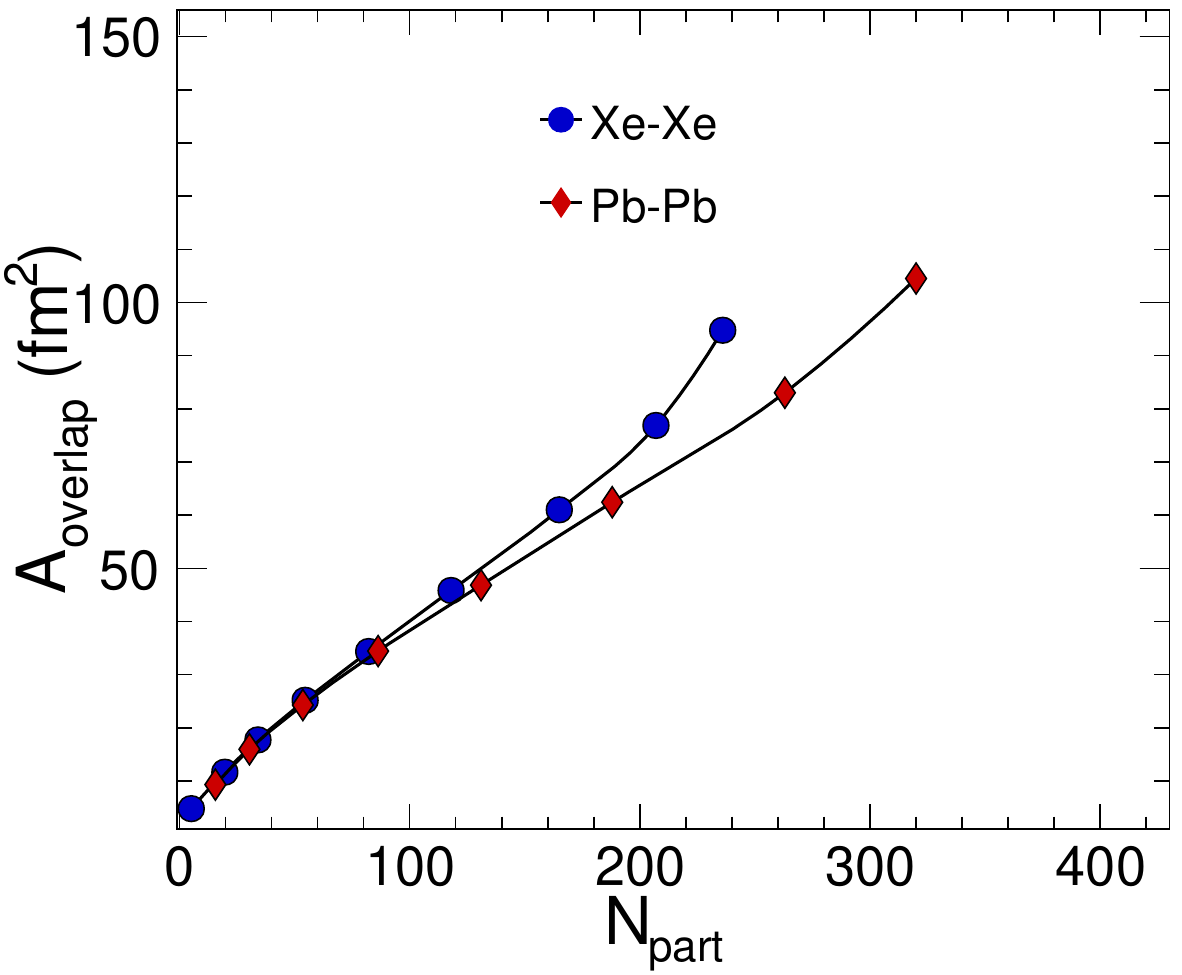}
    \caption{Generalized elliptic overlap area A$_{\text{overlap}}$ as a 
    function of the number of participating nucleons N$_{\rm part}$ for Xe-Xe 
collisions at \sqsn = 5.44 TeV and Pb-Pb collisions at \sqsn = 
5.02 TeV.}
    \label{area}
\end{figure}

The calculated overlap area is then investigated as a function of the number of
participating nucleons $N_{\text{part}}$ for Xe-Xe and Pb-Pb collisions, as shown in 
Fig.~\ref{area}. The overlap area increases monotonically with increase in 
$N_{\text{part}}$ for both systems. \par
In the peripheral region, the overlap areas of the two collision systems are 
comparable. However, with increasing $N_{\text{part}}$, a clear system-size 
dependence becomes evident. For mid-central and central collisions, the 
overlap area in Xe–Xe collisions exceeds that in Pb–Pb collisions at comparable 
$N_{\text{part}}$. This implies that Pb–Pb collisions can accommodate the same 
number of participating nucleons as Xe–Xe collisions within a more compact transverse 
geometry, resulting in a higher participant density in Pb–Pb collisions. The area of overlapping 
$A_{\text{overlap}}$ is an explicit input to the Bjorken initial energy density 
estimation. Consequently, these system size dependent differences in transverse 
geometry play a crucial role in the quantitative comparison of Bjorken initial energy 
densities, as discussed in the following sections.

\subsection{Estimation of Bjorken Initial Energy Density}\label{3B}
The Bjorken initial energy density is calculated for different centralities under 
two scenarios: a fixed formation time 
and varying $\tau_{0}$, using the 
Bjorken formula as given in Eq.~(\ref{eq:eneden}). The average transverse mass 
($\left<m_{T}\right>$) of pions is taken to be 0.562 GeV. The charged particle rapidity density 
$\left(\frac{dN_{ch}}{dy}\right)$ is taken from ALICE data~\cite{ALICE:2018cpu}. 

In the fixed-$\tau_0$ scenario, $\tau_0$ is set to 0.6 fm for all centrality 
classes. For Pb–Pb collisions at \sqsn = 5.02 
TeV, $\varepsilon_B$ is calculated for the fixed-$\tau_0$ case in different 
centrality classes. However, in the varying $\tau_0$ scenario, $\tau_0$ is extracted
using Eq.~(\ref{eq:eneden}) from the published energy-density values of Pb-Pb collisions
at \sqsn = 5.02 TeV for different centralities.  For Xe–Xe collisions at \sqsn = 5.44 TeV,
the Bjorken initial energy density ($\varepsilon_B$) is evaluated for both scenarios across
all centrality classes.
\subsection{Hydrodynamic Evolution Framework}\label{hy}
The space-time evolution of the QGP medium formed in Xe-Xe 
collisions at \sqsn = 5.44 TeV is studied within the framework of relativistic 
hydrodynamics. In this work, we focus on the longitudinal expansion of the system 
under the assumption of boost invariance in the transverse directions. We restrict 
our study to ideal and first-order viscous hydrodynamics.

The energy–momentum tensor for a locally thermalized relativistic viscous fluid is 
given by,
\begin{equation}\label{eq:T}
    T^{\alpha\beta} = \varepsilon u^\alpha u^\beta-P\Delta^{\alpha\beta} + \xi^{\alpha\beta} + \Delta^{\alpha\beta}\zeta\;\;,
\end{equation}

where $\varepsilon$ is energy density, $u^{\mu} = \left(\gamma,\; \gamma\; 
\vec{v}\right)$ is four-velocity 
of the fluid cell with $\gamma$ being the Lorentz factor, $P$ is the pressure. 
$\Delta^{\alpha\beta}=g^{\alpha\beta}-u^\alpha u^\beta$ is the projection operator 
with the metric tensor 
$g^{\alpha \beta}= \text{diag}\left(1, -1, -1, -1\right)$. $\xi^{\alpha \beta}$ represents shear viscous 
correction, and $\Delta^{\alpha\beta}\zeta$ corresponds to bulk viscous correction. At high temperature, 
the QGP medium is approximately conformal; therefore, the contribution from bulk viscosity is neglected 
throughout this study \cite{Weinberg:1972kfs}. In the absence of viscous correction terms, Eq.~(\ref{eq:T}) 
reduces to the ideal energy-momentum tensor.

In first-order hydrodynamic framework, shear stress tensor is expressed as,
\begin{equation}
    \xi^{\alpha\beta} = 2\eta \lambda^{\alpha\beta} \;,
\end{equation}
where $\eta$ is the shear viscosity, with
\begin{equation}
    \lambda^{\alpha\beta}= \frac{1}{2}\left(\nabla^\alpha u^\beta+ \nabla^\beta u^\alpha\right)-\frac{1}{3}\Delta^{\alpha\beta}\left(\nabla \cdot u\right),
\end{equation}
where $\nabla^\alpha = \partial^\alpha - u^\alpha u^\rho \partial_\rho$\; is the covariant derivative.\par 
The hydrodynamic evolution is governed by local energy-momentum conservation,
\begin{equation}\label{con}
    \partial_\alpha T^{\alpha\beta} = 0.
\end{equation}
To extract the energy evolution equation independent of momentum dynamics, we project Eq.~(\ref{con}) along the 
fluid velocity, i.e. $u_\beta   \partial_\alpha T^{\alpha\beta} = 0$. We then simplify this expression to obtain
the corresponding rate equations. For this purpose, we consider the ideal and viscous contributions separately.
\subsubsection{Ideal case}
     The ideal part of $u_\beta   \partial_\alpha T^{\alpha\beta}$ includes only the first two terms in 
     $T^{\alpha\beta}$, which gives 
    \begin{multline}\label{id}
        u_\beta \partial_{\alpha}\left[\varepsilon u^\alpha u^\beta-P\Delta^{\alpha\beta}\right]\\ = \partial_\alpha\left(\left(\varepsilon+P\right)u^\alpha\right)-u_\beta g^{\alpha\beta}\partial_\alpha P,
    \end{multline}
    which is obtained using $u_\beta u^\beta =1\;, \;u_\beta \left(\partial_\alpha u^\beta\right)=0\;,\; \text{and}\;\;u_\beta \Delta^{\alpha\beta} = 0$.\par
    In this study, we consider a simple $1+1$ dimensional case. The fluid has only longitudinal velocity, which 
    is assumed to be $v_z = \frac{z}{\tau}$ with $z$ and $\tau$ as longitudinal 
direction and proper time, respectively.\par
To simplify Eq.~(\ref{id}) further we use $\partial_\alpha u^\alpha = \frac{1}{\tau}$\; and $ u^\alpha \partial_\alpha = \frac{\partial}{\partial\tau}$\;,  which are derived explicitly in Appendix \ref{AB} and \ref{AC} respectively.
The simplification leads us to
    \begin{equation}
         u_\beta \partial_{\alpha}\left[\varepsilon u^\alpha u^\beta-P\Delta^{\alpha\beta}\right]\\= \frac{d \varepsilon}{d \tau} + \frac{\varepsilon+P}{\tau}.
    \end{equation}
   
  \subsubsection{Viscous case}
    In case of viscous contributions, the third term in $T^{\alpha \beta}$ contained in the expression 
    $u_\beta  \partial_\alpha T^{\alpha\beta}$ is considered. We have chosen Landau frame where the viscous 
    stress does not transport energy in the local rest frame. Consequently, the transversality of the shear 
    stress tensor implies $u_\beta \xi^{\alpha \beta} = 0$. The detailed derivation is explicitly shown in 
    Appendix \ref{AD}. Using this condition, we obtain,
    \begin{equation}
        u_\beta\partial_\alpha\xi^{\alpha\beta} = -\xi^{\alpha\beta}
         \partial_\alpha u_\beta = -2\eta \lambda^{\alpha \beta}\partial_\alpha u_\beta .
    \end{equation}
    Further simplification of this expression, explicitly derived in Appendix \ref{AE}, gives
    \begin{equation}
 u_\beta\partial_\alpha\xi^{\alpha\beta} = \frac{4\eta}{3\tau^2}.
    \end{equation}

By combining the ideal and viscous contributions, the rate equation for the energy density becomes
\begin{equation}\label{en}
    \frac{d \varepsilon}{d \tau} = -\frac{\varepsilon+P}{\tau} +  \frac{4\eta}{3\tau^2},
\end{equation}
which reproduces the ideal rate equation in the limit $\eta =0$. For the present study, we use a
conformal equation of state and the pressure, therefore becomes $P = \frac{\varepsilon}{3}$.

From relativistic kinetic theory, it has been observed that the shear viscosity primarily depends
on $\frac{\varepsilon}{T}$ \cite{Dutta:1999cn}. Hence, we can reasonably approximate $\eta$
as $ \eta \approx \kappa \frac{\varepsilon}{T}$, with a weakly temperature dependent
coefficient $\kappa$. In the subsequent calculations, $\kappa$ is treated as a constant parameter.
Now the rate equation for shear viscosity becomes
\begin{equation}\label{eta}
    \frac{d\eta}{d \tau} = \frac{\kappa}{T} \frac{d\varepsilon}{d\tau}\left(1-\frac{\varepsilon}{4aT^4}\right),
\end{equation}
where we have used the relation $\varepsilon = aT^4$ with the constant
$a =\frac{\pi^2}{30}\left(16+10.5N_f\right)$. In this study, we have considered the plasma
containing massless $u$, $d$, $s$ quarks, and number of quark flavors, $N_f=3$. Simplification
of Eq.~(\ref{eta}) leads to,
\begin{equation}\label{eta2}
    \frac{d\eta}{d\tau}=\frac{3\eta}{4\varepsilon}\frac{d\varepsilon}{d\tau}.
\end{equation}
Eqs.~(\ref{en}) and (\ref{eta2}) are two coupled differential equations. Both these equations are
simultaneously solved numerically to obtain the space-time evolution of the energy density and
shear viscosity.

\section{\label{D} Results}
\subsection{Estimation of Bjorken Initial Energy Density ($\varepsilon_{B}$)}
We have calculated the Bjorken initial energy density ($\varepsilon_{B}$) for 
different centrality classes in Xe-Xe collisions at \sqsn = 5.44 TeV for both fixed 
$\tau_{0}$ and varying $\tau_{0}$ cases. For Pb-Pb collisions at \sqsn = 5.02 TeV, the $\varepsilon_{B}$ values for different centrality classes are evaluated 
only for fixed $\tau_{0}$ case. In the fixed-$\tau_0$ case, $\tau_0$ is set to 0.6 fm for all the centrality classes. The area of overlap region 
A$_{\rm overlap}$ is calculated by using Eq.~(\ref{eq:overlap}) and along with the 
corresponding Bjorken initial energy density values obtained from Eq.~(\ref{eq:eneden}) 
for different centrality classes in Xe-Xe collisions at \sqsn = 5.44 TeV, are shown 
in Table~\ref{table1}. The Bjorken initial energy density ($\varepsilon_{B}$) values 
calculated for different centrality classes in Pb-Pb collisions at $\sqrt{s_{NN}} = 5.02$ TeV are shown in Table~\ref{table3}.

In the varying $\tau_{0}$ scenario, the formation time $\tau_{0}$ for different 
centrality classes is extracted from the published energy density ($\varepsilon_{B}$) 
values for Pb-Pb collisions at \sqsn = 5.02 TeV using Eq.~(\ref{eq:eneden}). The 
extracted $\tau_{0}$ values, along with the corresponding area of overlap region 
A$_{\rm overlap}$ calculated by using Eq.~(\ref{eq:overlap}) and the published Bjorken 
initial energy density ($\varepsilon_{B}$) values from ALICE data for different 
centrality classes in Pb-Pb collisions at \sqsn = 5.02 TeV are shown in Table~\ref{table5}. The area of overlap region 
A$_{\rm overlap}$ and Bjorken initial energy density ($\varepsilon_{B}$) values  for different centrality classes are 
calculated in Xe-Xe collisions at \sqsn = 5.44 TeV by using the $\tau_{0}$ extracted from 
published energy density ($\varepsilon_{B}$) values of Pb-Pb collisions at 
$\sqrt{s_{NN}} = 5.02$ TeV, and are shown in 
Table~\ref{table6}.\par
\begin{figure}[htb]
    \centering
    \includegraphics[width=1\linewidth]{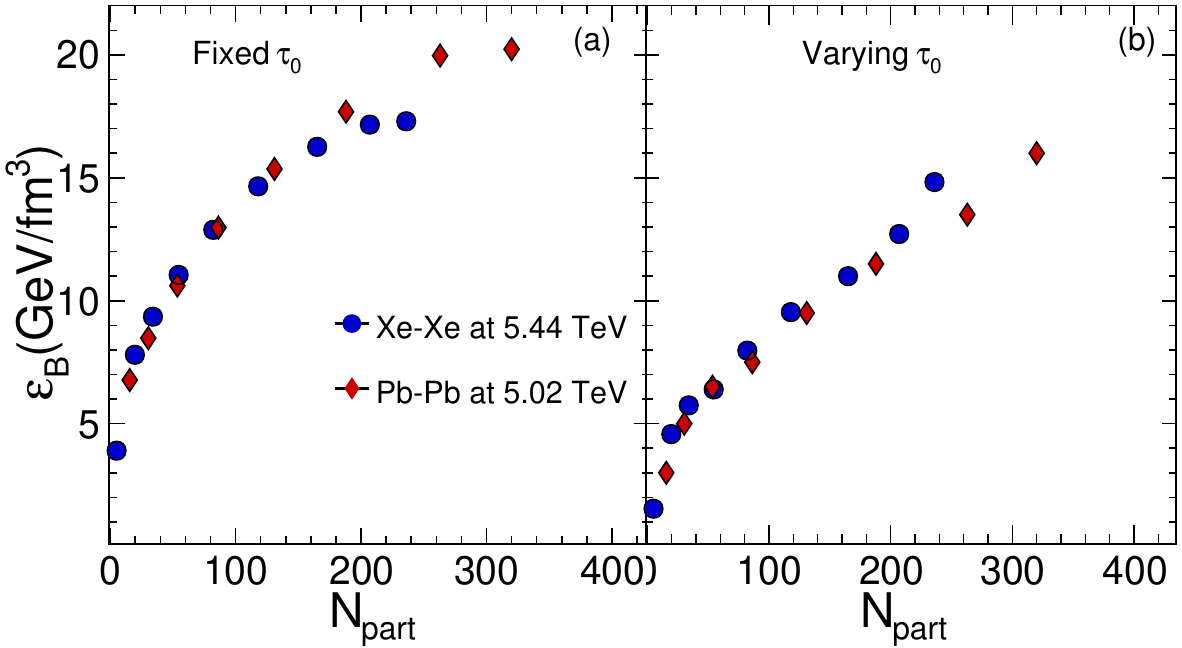}
    \caption{Bjorken initial energy density $\varepsilon_B$ as a function of N$_{\text{part}}$ for
      Xe-Xe and Pb-Pb collisions under (a) fixed formation time $\tau_0$ and (b) centrality dependent
      $\tau_0$ scenarios.}
    \label{cmp_var}
\end{figure}

The variation of the Bjorken initial energy density ($\varepsilon_{B}$) with the 
number of participant nucleons $(N_{\rm part})$ for both 
the fixed $\tau_{0}$ and varying $\tau_{0}$ cases in Xe-Xe at \sqsn = 5.44 TeV and 
Pb-Pb collisions at \sqsn = 5.02 TeV is shown in Fig.~\ref{cmp_var}.
The Bjorken initial energy density increases monotonically with N$_{\rm part}$ for both of the colliding systems. This behavior is a direct consequence of the growth of the transverse overlap area and particle production with increase in collision centrality. The varying-$\tau_0$ scenario leads to comparatively lower $\varepsilon_B$ values, especially in peripheral collisions. This behavior highlights the role of a centrality-dependent formation time in moderating the initial energy density, while preserving the overall N$_{\rm part}$ scaling.
\begin{table*}
	\caption{\label{table1} Bjorken initial energy density $(\varepsilon_{B})$ calculated by using the 
    charged particle rapidity density $\left(\frac{dN_{ch}}{dy}\right)$~\cite{ALICE:2018cpu}, area of overlap region
    $(A_{\rm overlap})$, the formation time $\tau_{0}$ as given in this table for 
different centrality classes in 
    Xe-Xe collisions at \sqsn = 5.44 TeV.}
	\begin{ruledtabular}
		\begin{tabular}{ccccccc}
			&\multicolumn{0}{c}{}\\
			Centrality class (\%)&$N_{\rm part}$&$\frac{dN_{ch}}{d\eta}$&b 
(fm)&$A_{\rm overlap, Cal}$ (fm)$^{2}$&$\tau_{0}$ (fm)&$\varepsilon_{\rm B, Cal}$ 
($GeV/(fm)^{3}$)\\ \hline
			0-5&236&1167&1.817&94.778&0.600&17.299\\
			5-10&207&939&3.320&76.868&0.600&17.163\\
			10-20&165&706&4.698&61.004&0.600&16.260\\
			20-30&118&478&6.086&45.830&0.600&14.653\\
			30-40&82.2&315&7.209&34.352&0.600&12.883\\
			40-50&54.6&198&8.179&25.178&0.600&11.049\\
			50-60&34.1&118&9.042&17.727&0.600&9.352\\
			60-70&19.7&64.7&9.829&11.650&0.600&7.802\\
			70-90&5.13&13.3&10.900&4.790&0.600&3.901\\
		\end{tabular}
	\end{ruledtabular}
\end{table*}

\begin{table*}
	\caption{\label{table3} Bjorken initial energy density $(\varepsilon_{B})$ 
	calculated by using the charged particle rapidity density 
$\left(\frac{dN_{ch}}{dy}\right)$~\cite{ALICE:2018cpu}, area of overlap region 
$(A_{\rm overlap})$ and the formation time $\tau_{0}$ as given in this table for 
different centrality classes in Pb-Pb collisions at \sqsn = 5.02 TeV.}
	\begin{ruledtabular}
		\begin{tabular}{ccccccc}
			&\multicolumn{0}{c}{}\\
			Centrality class (\%)&$N_{\rm part}$&$\frac{dN_{ch}}{d\eta}$&b 
(fm)&$A_{\rm Overlap, Cal} (fm)^{2}$&$\tau_{0} (fm)$&$\varepsilon_{\rm B, Cal}$ 
($GeV/(fm)^{3}$)\\ \hline
			0-10&320&1505&3.867&104.510&0.600&20.232\\
			10-20&263&1180&5.468&83.019&0.600&19.970\\
			20-30&188&786&7.083&62.429&0.600&17.689\\
			30-40&131&512&8.391&46.834&0.600&15.360\\
			40-50&86.3&318&9.514&34.436&0.600&12.974\\
			50-60&53.6&183&10.521&24.281&0.600&10.616\\
			60-70&30.4&96.3&11.443&15.961&0.600&8.477\\
			70-80&15.6&44.9&12.293&9.314&0.600&6.773\\
		\end{tabular}
	\end{ruledtabular}
\end{table*}

\begin{table*}
	\caption{\label{table5} The formation time $\tau_{0}$ calculated by using the charged particle rapidity density $\left(\frac{dN_{ch}}{dy}\right)$~\cite{ALICE:2018cpu}, area of overlap region $(A_{overlap})$ and the Bjorken initial energy density $(\varepsilon_{B})$ as given in this table for different centrality classes in Pb-Pb collisions at $\sqrt{s_{NN}} = 5.02$ TeV~\cite{Prasad:2021bdq}.}
	\begin{ruledtabular}
		\begin{tabular}{ccccccc}
			&\multicolumn{0}{c}{}\\
		  Centrality class (\%)&$N_{part}$&$\frac{dN_{ch}}{d\eta}$&b
                  (fm)&$A_{Overlap, Cal} (fm)^{2}$&$\tau_{0, Cal} (fm)$&$\varepsilon_{B, Pub}$
                  ($GeV/(fm)^{3}$)\\ \hline
			0-10&320&1505&3.867&104.510&0.758&16.000\\
			10-20&263&1180&5.468&83.019&0.887&13.500\\
			20-30&188&786&7.083&62.429&0.922&11.500\\
			30-40&131&512&8.391&46.834&0.970&9.500\\
			40-50&86.3&318&9.514&34.436&1.037&7.500\\
			50-60&53.6&183&10.521&24.281&0.977&6.500\\
			60-70&30.4&96.3&11.443&15.961&1.023&5.000\\
			70-80&15.6&44.9&12.293&9.314&1.354&3.000\\
		\end{tabular}
	\end{ruledtabular}
\end{table*}
\begin{table*}
  \caption{\label{table6} Bjorken initial energy density $(\varepsilon_{B})$ calculated by using
    the charged particle rapidity density $\left(\frac{dN_{ch}}{dy}\right)$~\cite{ALICE:2018cpu}, area of
    overlap region $(A_{overlap})$, the formation time $\tau_{0}$ as given in this table for
    different centrality classes in Xe-Xe collisions at $\sqrt{s_{NN}} = 5.44$ TeV.}
	\begin{ruledtabular}
		\begin{tabular}{ccccccc}
			&\multicolumn{0}{c}{}\\
		  Centrality class (\%)&$N_{part}$&$\frac{dN_{ch}}{d\eta}$&b
                  (fm)&$A_{overlap, Cal}$ (fm)$^{2}$&$\tau_{0, Cal}$ (fm)&$\varepsilon_{B, Cal}$ ($GeV/(fm)^{3}$)\\ \hline
			0-5&236&1167&1.817&94.778&0.700&14.828\\
			5-10&207&939&3.320&76.868&0.810&12.713\\
			10-20&165&706&4.698&61.004&0.887&10.999\\
			20-30&118&478&6.086&45.830&0.922&9.536\\
			30-40&82.2&315&7.209&34.352&0.970&7.969\\
			40-50&54.6&198&8.179&25.178&1.037&6.393\\
			50-60&34.1&118&9.042&17.727&0.977&5.743\\
			60-70&19.7&64.7&9.829&11.650&1.023&4.576\\
			70-90&5.13&13.3&10.900&4.790&1.519&1.541\\
		\end{tabular}
	\end{ruledtabular}
\end{table*}

\begin{figure}[htb]
    \centering
    \includegraphics[width=1\linewidth]{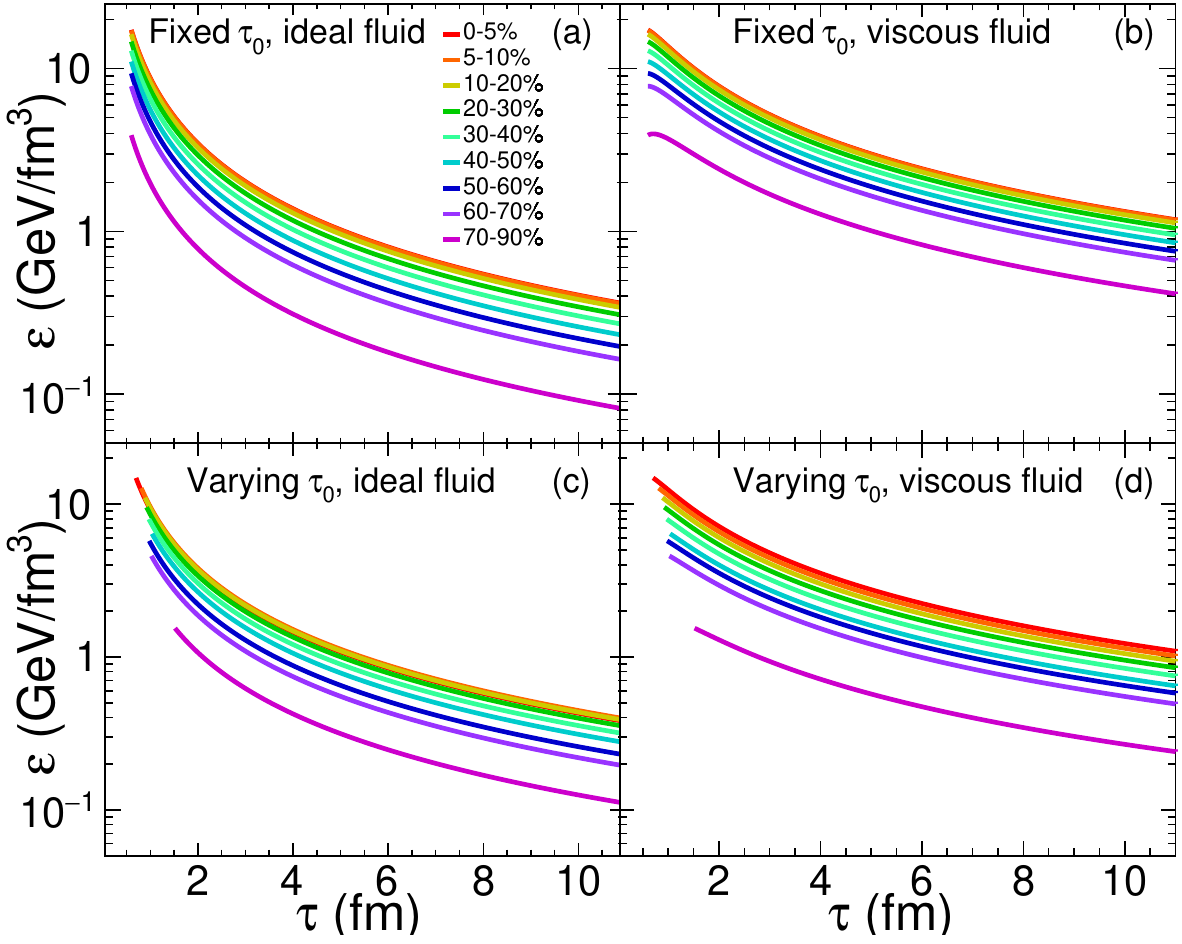}
    \caption{Time evolution of energy density $\varepsilon(\tau)$ for different 
centrality classes in Xe-Xe collisions at \sqsn = 5.44 TeV for ideal and 
viscous hydrodynamics under fixed and varying $\tau_0$ scenarios.}
    \label{en_ev_var}
\end{figure}

\subsection{Evolution of the QGP Medium}
Using the Bjorken initial energy density ($\varepsilon_B$) and the corresponding 
formation time $\tau_0$ as initial conditions, the subsequent space-time evolution of 
the QGP medium formed in Xe-Xe collisions at \sqsn = 5.44 TeV is studied by 
numerically solving the coupled hydrodynamic equations derived in Sec.~\ref{hy}. 
Both ideal and first-order viscous hydrodynamic evolution are studied under the 
assumption of boost-invariant longitudinal expansion. 

\subsubsection{Longitudinal Evolution of Energy Density ($\varepsilon$)}
Figure~\ref{en_ev_var} shows the evolution of energy density for different 
centrality classes under two distinct scenarios:$(i)$ varying $\tau_0$, where the 
formation time increases from central to peripheral collisions as extracted from 
Pb-Pb data at \sqsn = 5.02 TeV, and $(ii)$ a fixed $\tau_0$, where a constant value 
$\tau_0$ = 0.6 fm is used for all centralities to isolate the role of initial time 
as a control parameter. In both scenarios, the energy density decreases monotonically 
with proper time, reflecting the longitudinal work performed by the expanding medium 
as the Co-moving volume increases. In the ideal case, the evolution follows Bjorken 
scaling  $\varepsilon \propto \tau^{-4/3}$ for conformal equation of state. With the 
inclusion of viscous effects, the dilution of energy density is reduced, leading to 
systematically higher energy densities at later times compared to the ideal case. 
This behavior arises from viscous contributions to the longitudinal pressure, which 
counteract the rapid dilution induced by longitudinal expansion.

To disentangle the effect of formation time $\tau_0$ from that of the initial energy 
density $\varepsilon_B$, a fixed $\tau_0$ scenario is considered as shown in Figs.~\ref{en_ev_var}(a, b). Although, this assumption is not realistic, it is considered 
to investigate the differences between centrality classes arising solely from 
variations in $\varepsilon_B$, which are determined by collision geometry and particle 
production in the collisions.

Figures~\ref{en_ev_var}(c, d) show evolution of energy density in varying $\tau_0$ scenario. In peripheral 
collisions, the hydrodynamic evolution begins at larger $\tau_0$ with smaller 
initial energy density $\varepsilon_B$. Thus, these systems reach dilute conditions more 
rapidly and exhibit a shorter lifetime of the dense QGP phase. In contrast, central 
collisions are characterized by smaller $\tau_0$ and larger $\varepsilon_B$, allowing 
the system to remain in the de-confined phase for a longer duration. The higher 
participant densities in central collisions favor faster local equilibrium, thereby 
justifying the earlier applicability of hydrodynamics. Comparison of the fixed and varying $\tau_0$ cases 
indicates that a centrality dependent formation time enhances the contrast between 
central and peripheral collisions, particularly during the early stages of the 
evolution. 
\begin{figure}[htb]
    \centering
    \includegraphics[width=1\linewidth]{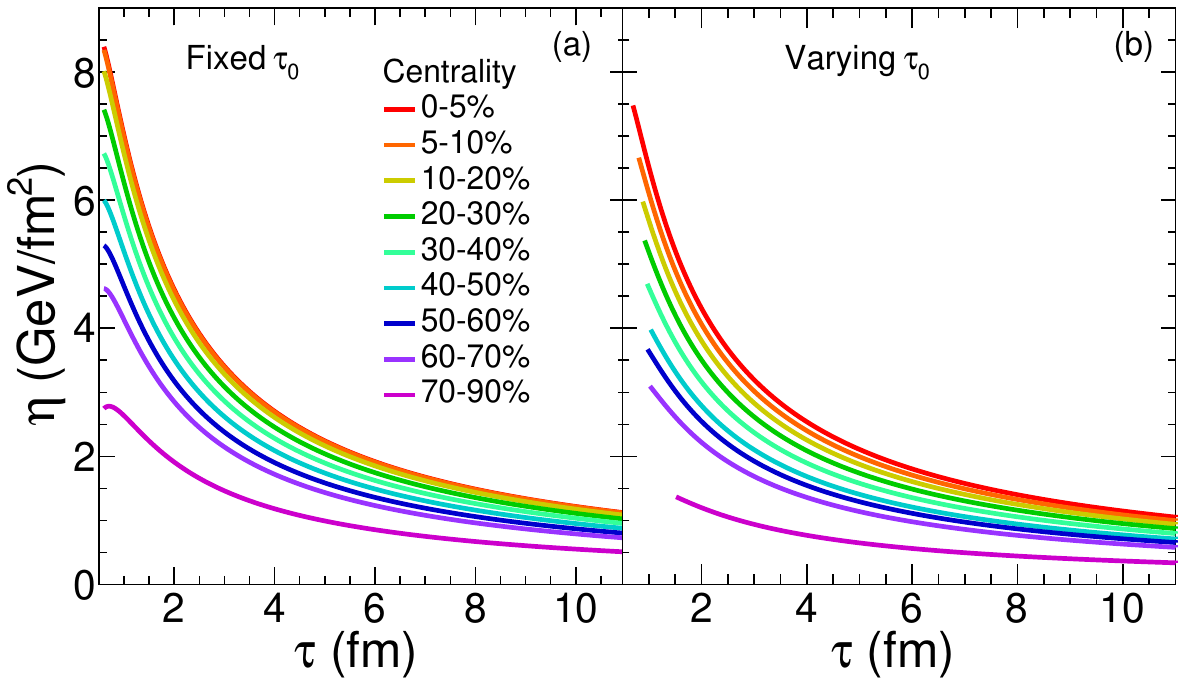}
    \caption{Time evolution of shear viscosity $\eta(\tau)$ for different centrality 
classes in Xe-Xe collisions for (a) fixed $\tau_0$ and (b) centrality dependent 
$\tau_0$ scenarios.}
    \label{eta_ev_var}
\end{figure}

\subsubsection{Evolution of Shear Viscosity ($\eta$)}
Figure~\ref{eta_ev_var} shows proper time evolution of the shear viscosity 
$\eta\left(\tau\right)$ for different centrality classes for both fixed and varying 
$\tau_0$ scenarios. In all cases, $\eta$ decreases monotonically with proper time 
($\tau$) reflecting the cooling of the system. Due to the longitudinal expansion of 
the medium, the temperature and energy density decrease with time, leading to a corresponding 
reduction in the magnitude of shear viscosity. Central collisions exhibit larger 
initial values of $\eta$ and therefore sustain higher viscosity over a longer 
duration as compared to peripheral collisions, consistent with their larger 
$\varepsilon_B$.  

Evolution of viscosity in fixed $\tau_0$ case is shown in Fig.~\ref{eta_ev_var}(a) where the ordering of $\eta\left(\tau\right)$ across centrality classes 
is determined solely by collision geometry and particle production. The corresponding case of varying $\tau_0$ is shown in Fig.~\ref {eta_ev_var}(b). Peripheral collisions
have smaller initial $\eta$ values due to their significantly lower $\varepsilon_B$ 
values causing viscous effects to become sub-dominant at relatively early times. In 
contrast, central collisions retain substantial viscous contributions during the 
early stages of evolution, which plays a crucial role in moderating the cooling rate 
and entropy generation, as discussed later in this study. 

The evolution of $\varepsilon$ and $\eta$ gives a coherent description of the 
longitudinal dynamics of the QGP medium. In the following sections, we explore the 
implications of this evolution for entropy production and the effective lifetime of 
the disordered de-confined QGP medium.
\begin{figure}[htb]
    \centering
    \includegraphics[width=1\linewidth]{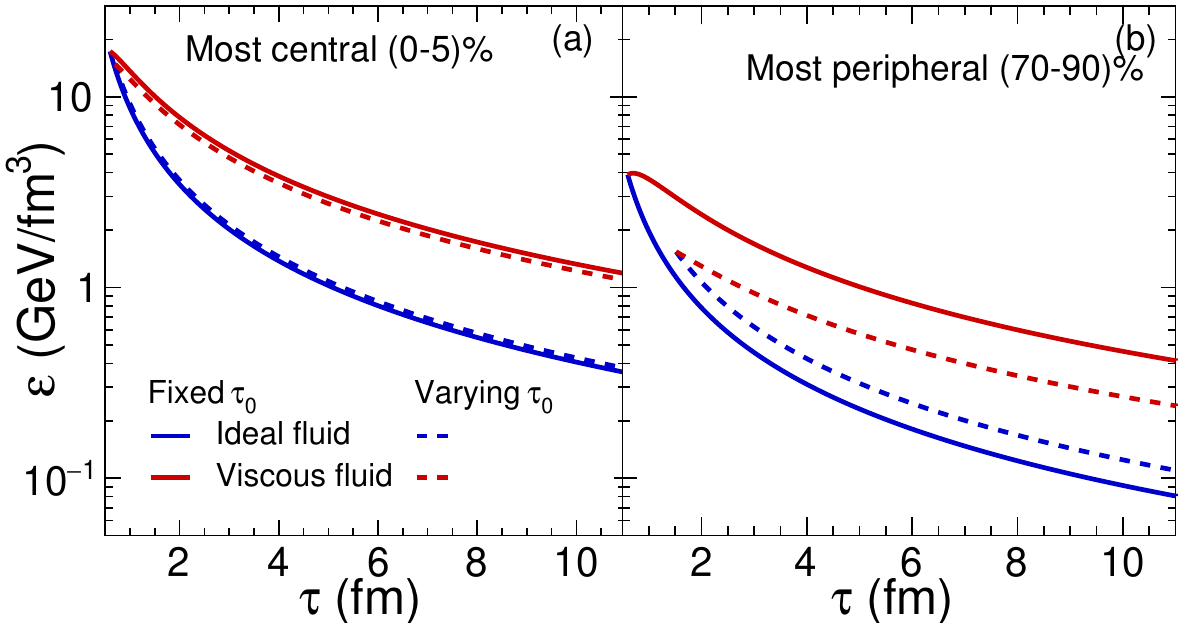}
    \caption{Comparison of ideal and viscous energy density evolution for the most 
central $(0-5 \%)$ and most peripheral $(70-90 \%)$ Xe-Xe collisions under fixed and 
varying $\tau_0$ assumptions.}
    \label{comp}
\end{figure}

To further elucidate the effect of shear viscosity on the dynamical evolution, a 
direct comparison between ideal and viscous fluid dynamics is performed for the most 
central $\left(0-5\%\right)$ and most peripheral $\left(70-90\%\right)$ centrality 
classes in Xe-Xe collisions. Shear viscosity contributes to the longitudinal 
pressure, which partially counterbalances the work done by the system during 
longitudinal expansion. As a result, the inclusion of shear viscosity leads to a 
systematically slower cooling of the medium compared to the ideal case, in both 
central and peripheral collisions. This is consistent for both fixed and 
centrality-dependent formation-time scenarios as shown in Fig.~\ref{comp}.

A clear distinction emerges in the most peripheral collisions, where viscous 
effects are significantly more pronounced in the fixed $\tau_0$ case than in the 
varying $\tau_0$ scenario. Upon fixing $\tau_0$ for all centralities we force 
peripheral collisions to undergo local equilibrium at an early proper time despite 
their lower $\varepsilon_B$ value. At such early times the longitudinal expansion rate 
is large, which enhances the relative contribution of the viscous terms to the 
evolution equations and resulting in an artificially strong viscous correction in 
the most peripheral collisions. However, in the case of varying $\tau_0$, onset of 
evolution happens at larger formation time. This delay reduces the effective 
expansion rate and suppresses viscous corrections, leading to a much smaller 
deviation between ideal and viscous evolution, as shown in Fig. \ref{comp}(b). 
This behavior is expected, as dilute systems are supposed to thermalize later and 
undergo a shorter viscous hydrodynamic phase.

In the most central collisions, as shown in Fig. \ref{comp}(a), the effect of 
viscosity is more pronounced. Here as well, fixing $\tau_0$ tends to overestimate 
the viscous corrections as compared to the more realistic centrality-dependent 
$\tau_0$ case. However, due to smaller $\tau_0$ and higher $\varepsilon_B$ values in 
central collisions for both the cases, the quantitative differences between fixed 
and varying $\tau_0$ cases remains small.
\subsubsection{Entropy production in Longitudinal Expansion}
\begin{figure}[htb]
    \centering
    \includegraphics[width=1\linewidth]{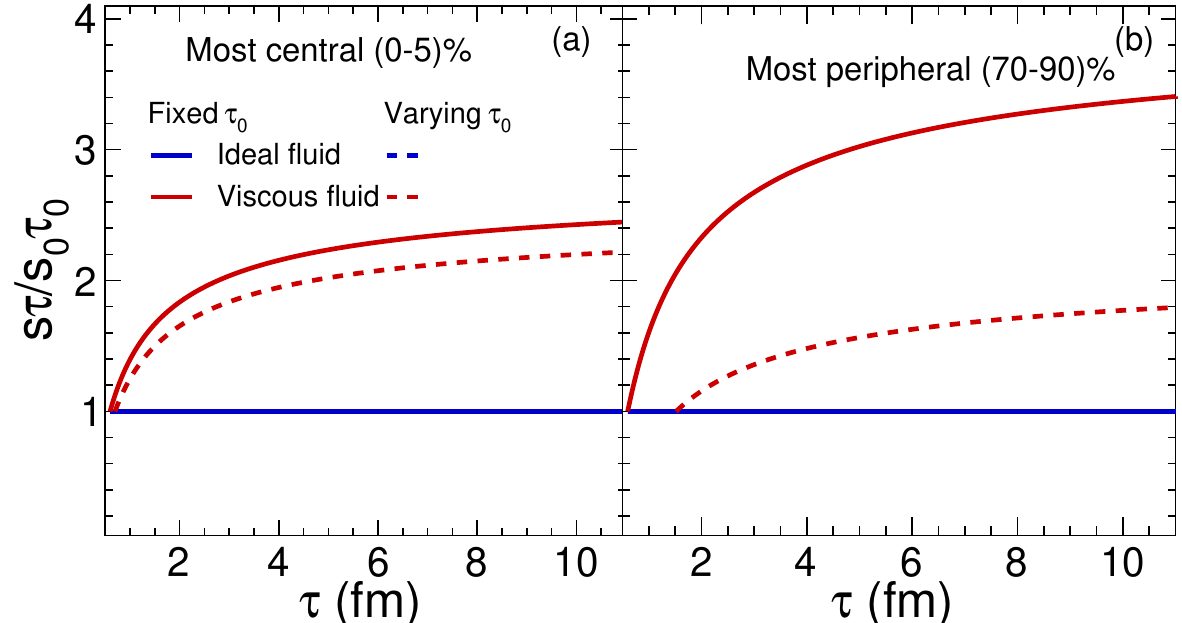}
    \caption{Entropy production quantified through dimensionless $s(\tau)\tau /s(\tau_0)\tau_0$
      for ideal and viscous evolution in central and peripheral Xe-Xe collisions.}
    \label{entropy}
\end{figure}

The same mechanism is reflected in the entropy production during evolution, as 
shown in Fig.~\ref{entropy}. For an ideal fluid undergoing boost-invariant 
longitudinal expansion, entropy per unit rapidity is conserved. Therefore, the ratio 
$s\left(\tau\right)\tau/s\left(\tau_0\right)\tau_0$ remains constant throughout the 
evolution for different cases of ideal fluid. In contrast, the viscous evolution 
exhibits a monotonic increase of this ratio, reflecting entropy production due to 
shear viscosity. This entropy generation is most significant at the early stages of 
the evolution, when velocity gradients are largest. Then the ratio gradually 
saturates at later times as the medium cools and the expansion rate decreases. 
A noteworthy observation from Fig.~\ref{entropy} is that, in the viscous case, 
the centrality-dependent $\tau_0$ scenario results in smaller total entropy 
production as compared to fixed $\tau_0$ case in both central and peripheral 
collisions. This behavior is a direct consequence of the delayed onset of 
hydrodynamic evolution in varying $\tau_0$ case, particularly for peripheral 
collisions. Since entropy production is dominated by early stages of evolution, 
initiating the hydrodynamics at a larger formation time effectively reduces the 
interval over which dissipative processes act, thereby leads to a smaller increase 
in entropy. In contrast, the fixed $\tau_0$ case systematically overestimates the 
entropy production. This overestimation is more pronounced in the most peripheral 
collisions, where the medium is forced to start undergoing evolution at earlier 
formation time with low initial energy density. These observations establish a clear 
correlation between the slower energy density evolution in viscous hydrodynamics and 
the associated entropy production.

\subsubsection{Lifetime of the QGP Medium}
The hydrodynamic evolution discussed in the preceding sections allows us to 
estimate the effective lifetime of the QGP medium formed in Xe-Xe collisions at 
\sqsn = 5.44 TeV. Within the Bjorken framework, the QGP phase is assumed to persist 
as long as the energy density remains above a critical value. Following standard 
phenomenological practice, we take the critical energy density as 
$\varepsilon_c = 1$ GeV/fm$^3$ for the onset of hadronization. The proper time at 
which $\varepsilon$ drops below this threshold is defined as lifetime of QGP, 
denoted by $\tau_{\text{critical}}$. 

\begin{figure}[htb]
    \centering
    \includegraphics[width=1\linewidth]{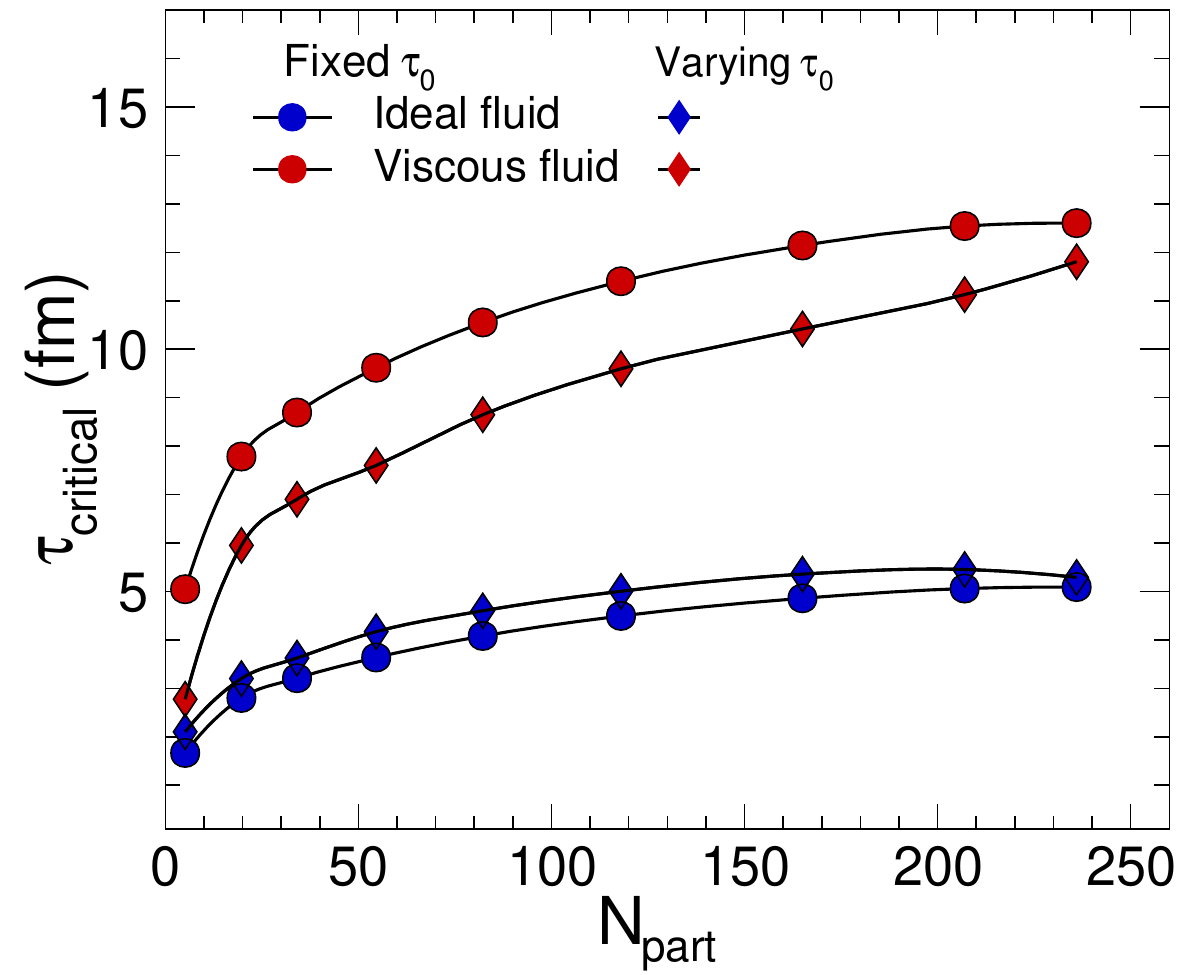}
    \caption{Lifetime of the QGP phase $\tau_{\text{critical}}$ (defined by 
    $\varepsilon = 1$ GeV/fm$^3$) as a function of N$_{\text{part}}$ for ideal and 
viscous evolution under fixed and varying $\tau_0$ scenarios.}
    \label{lifetime_var}
\end{figure}

Figure~\ref{lifetime_var} shows the QGP lifetime as a function of the number of 
participating nucleons $N_{\rm part}$, for both ideal and viscous case for both 
varying and fixed $\tau_0$ scenarios.  The lifetime increases monotonically with 
increasing $N_{\text{part}}$. Central collisions, characterized by larger overlap 
area and higher $\varepsilon_B$, remain longer in the de-confined phase whereas 
peripheral collisions cool more rapidly and undergo earlier hadronization. This trend 
reflects stronger longitudinal dilution exhibited by smaller and more dilute systems 
and provides a consistency check of the Bjorken initial conditions and subsequent 
evolution. As discussed earlier, shear viscosity generates additional longitudinal 
pressure, which slows the cooling rate of the medium. Consequently, $\varepsilon$ 
remains above $1$ GeV/fm$^3$ for a longer time, leading to an increased lifetime of 
QGP.

The distinction between fixed and varying $\tau_0$ cases is relatively small for 
ideal fluid. However, upon including viscosity the difference becomes substantial, 
particularly in peripheral collisions. This behavior can be understood from the 
structure of the hydrodynamic evolution equations in conjunction with the Bjorken 
initial condition. In ideal Bjorken flow, lifetime  of the QGP exhibits a weak 
dependence on the formation time, scaling as $\tau_{\rm critical} \propto 
\tau_0^{1/4}$. In contrast, the viscous term in the evolution equation scales as $1/\tau^2$, rendering the evolution 
sensitive to early-time dynamics. As a result, memory effects from the early stages 
strongly influence the extracted QGP lifetime in presence of viscosity.

In peripheral collisions, fixing $\tau_0$ leads to an overestimation of the QGP lifetime by forcing the system to undergo early evolution with a reduced $\varepsilon_B$.

Upon employing a centrality-dependent formation time delays the onset of hydrodynamic 
evolution. This delayed onset suppresses early time viscous contributions, resulting 
in a much shorter and more realistic QGP lifetime. For the most central collisions, 
the values of $\tau_0$ are small and comparable in both scenarios. Viscous effects 
act over a similar early time interval, thus the distinction between fixed and 
varying $\tau_0$ scenarios remains modest. 

\subsubsection{Comparison of $\varepsilon$ Evolution in Xe-Xe and Pb-Pb Collisions}
In this subsection, we compare the hydrodynamic evolution of the QGP medium formed 
in Xe-Xe collisions at \sqsn = 5.44 TeV with that in Pb-Pb collisions at \sqsn = 5.02 TeV. This study investigates the dependence of the evolution on the 
colliding system and the collision energy within the Bjorken framework.  
\begin{figure}[htb]
    \centering
\includegraphics[width=1\linewidth]{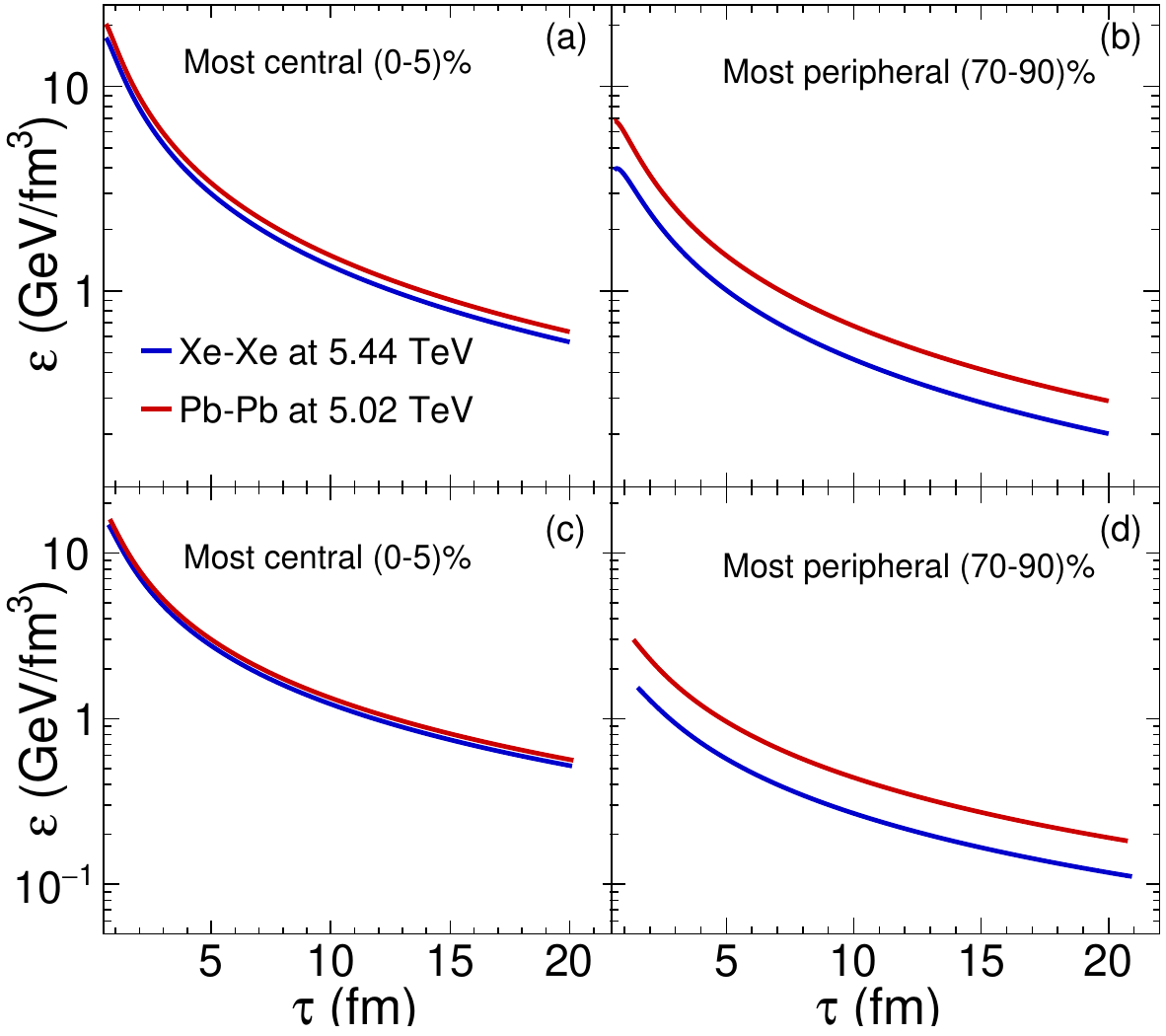}
    \caption{Comparison of longitudinal energy density evolution in Xe-Xe and Pb-Pb
      collisions for most central and peripheral events. Top panels show fixed $\tau_0$ case and bottom panels show the varying $\tau_0$ case.}
    \label{xe_pb_var}
\end{figure}

For the most central collisions, as shown in Figs. \ref{xe_pb_var}(a, c), the energy density evolution in Pb-Pb collisions remains systematically higher than that in Xe-Xe collisions for both fixed and varying $\tau_0$ cases. This difference is attributed to the ratio of charged particle multiplicity to the effective Bjorken volume at the onset of the hydrodynamic evolution, $\left(\left(dN_{\text{ch}}/dy\right)/\left(A_{\text{overlap}}\tau_0\right)\right)$. The ratio value for Pb-Pb collisions is higher than that in Xe-Xe collisions, causing higher $\varepsilon$ for the former. 
Figures~\ref{xe_pb_var}(b, d) show the ordering observed for the most peripheral collisions. Here the separation between the two systems becomes more pronounced. This again depicts a direct correlation with the ordering of the multiplicity to volume ratio.
Due to the larger relative differences among systems and energies, peripheral collisions
provide a more sensitive environment for disentangling system-size and collisions-energy
dependent effects in the longitudinal hydrodynamic evolution. 

\section{\label{E} Summary}
We study the Bjorken initial energy density ($\varepsilon_B$) in Xe-Xe 
collisions at \sqsn = 5.44 TeV using the Bjorken prescription. To 
quantify the role of formation time, both fixed and centrality-dependent $\tau_0$ 
cases are considered. The centrality dependence of the $\varepsilon_B$ is 
studied using a generalized and more realistic elliptic transverse overlap geometry, 
extending the conventional circular approximation to mid-central and peripheral 
collisions. Using the initial energy density ($\varepsilon_B$) and formation time 
($\tau_0$), viscous longitudinal hydrodynamic evolution of the system formed in heavy-ion 
collisions is examined. We find that the formation time strongly influences the 
magnitude and centrality dependence of the initial energy density, with particularly 
pronounced effects in peripheral collisions. The inclusion of shear viscosity slows the longitudinal expansion and enhances entropy 
production. These viscous effects, dominated by early-time dynamics, are amplified 
when a fixed formation time is imposed, especially in peripheral collisions. In contrast, a centrality-dependent formation time reduces both viscous corrections and entropy 
production by delaying the onset of hydrodynamic evolution in dilute systems. 
 The lifetime of quark-gluon plasma ($\tau_{\rm critical}$) increases with centrality 
and is further  extended by the inclusion of viscosity in the model. To investigate the system-size dependence, results from Xe-Xe collisions are compared with those
from Pb-Pb collisions at \sqsn = 5.02 TeV. For the most central 
collisions, the energy densities and their time evolution are found to be nearly identical 
for both Xe-Xe and Pb-Pb collisions. In contrast, peripheral Pb-Pb collisions exhibit systematically higher energy densities compared to peripheral Xe-Xe 
collisions.

\nocite{*}

\appendix
\section{Overlapping Area Calculation}\label{AA}
The transverse overlapping area is equal to the sum of the areas of two circular segments, corresponding to participating nuclei.\par
\begin{figure}[htb]
    \centering
    \includegraphics[width=0.8\linewidth]{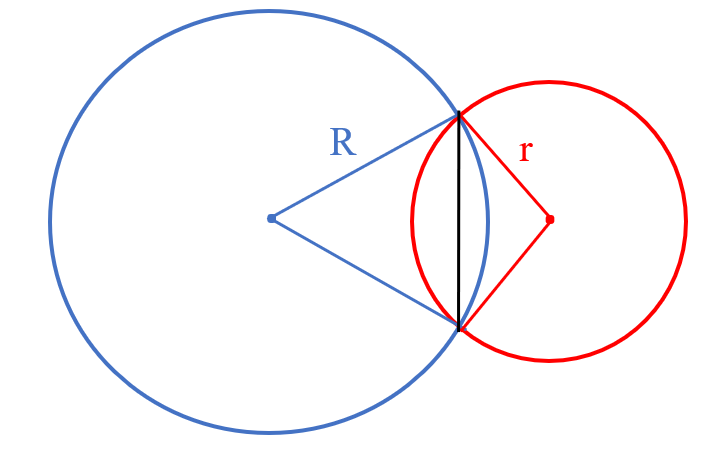}
    \label{fig:placeholder}
\end{figure}
The area of the circular segment of a nucleus can be found in the following way.
\begin{figure}[htb]
    \centering
    \includegraphics[width=0.65\linewidth]{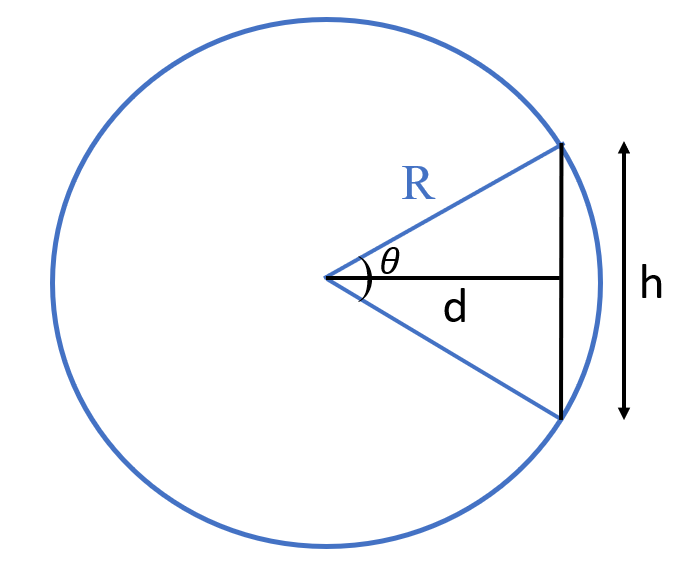}
    \label{segment}
\end{figure}

If we subtract the area of the triangle from the area of the circular sector, we will get the area of the circular segment corresponding to the nucleus of radius $R$, $A_{cs}^R$.
\begin{equation}
    A_{cs}^R = \frac{1}{2}R^2\theta- \frac{1}{2}dh .
\end{equation}
Using few simple trigonometric tricks, the last expression is simplified as,
\begin{equation}
    A_{cs}^R = R^2\cos^{-1}\left(\frac{d}{R}\right)-d\sqrt{R^2-d^2}.
\end{equation}

Similarly, area of circular segment corresponding to nucleus of radius $r$, $A_{cs}^r$ is written as,
\begin{equation}
    A_{cs}^r = r^2\cos^{-1}\left(\frac{b-d}{r}\right)-\left(b-d\right)\sqrt{r^2-\left(b-d\right)^2}.
\end{equation}
\begin{figure}[htb]
    \centering
    \includegraphics[width=0.8\linewidth]{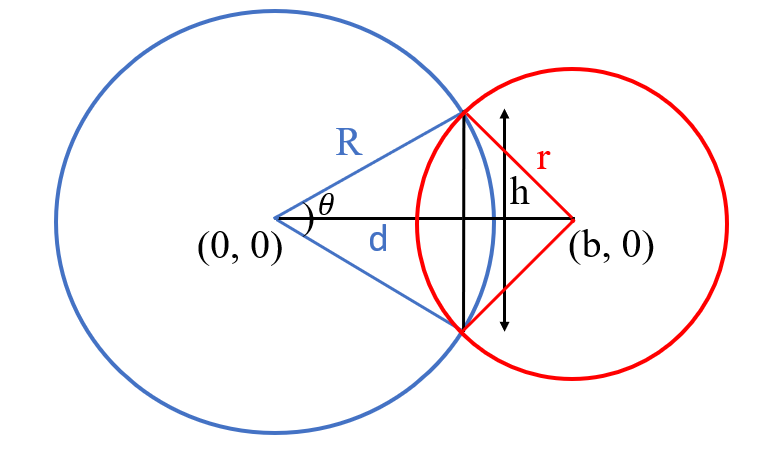}
    \label{fig:placeholder}
\end{figure}
Now equation of two circular nuclei can be written as,
\begin{equation}\label{A4}
    x^2+y^2 = R^2\;, \;\;\text{and} \;\;\;\left(x-b\right)^2+y^2 = r^2.
\end{equation}
Solving these two equations we get,
\begin{multline}
        x = \frac{b^2-r^2+R^2}{2b} \;,\\ y = \frac{1}{2b}\sqrt{\left(b+R+r\right)\left(b+R-r\right)\left(r+b-R\right)\left(r-b+R\right)}. 
\end{multline}

It is easy to recognize that $x=d$ and $y =\frac{h}{2}$. Now the overlapping area can be found as,
\begin{equation}
    A_{\text{overlap}} =A_{cs}^R + A_{cs}^r ,
\end{equation}
\begin{multline*}
        A_{\text{overlap}} = R^2 \cos^{-1} \left(\frac{x}{R}\right) + r^2 \cos^{-1} \left(\frac{b-x}{r}\right)\\- x\sqrt{R^2-x^2}-\left(b-x\right)\sqrt{r^2-\left(b-x\right)^2}.
\end{multline*}
Using Eq.~(\ref{A4}), the last two terms in the expression can be simplified to give $-by$. So the final expression of the overlapping area becomes,
\begin{multline}
    A_{\text{overlap}} =\; R^2 \cos^{-1} \left(\frac{b^2-r^2+R^2}{2bR}\right) \\+ r^2 \cos^{-1} \left(\frac{b^2+r^2-R^2}{2br}\right) \\
    -\frac{1}{2} \sqrt{\left(b+r+R\right)\left(b-r+R\right)\left(r+b-R\right)\left(r-b+R\right)}.
\end{multline}
\section{Calculation of $\partial_\alpha u^\alpha$}\label{AB}
Milne coordinates provide a more physical efficient framework to describe the boost-invariant rapid longitudinal expansion of the medium formed in heavy-ion collisions. Calculations get simpler by allowing the system to evolve in proper time with simple flow fields and conservation equations. Hence for all the subsequent calculations, we will consider Milne coordinates given by $\tau = \sqrt{t^2-z^2}\;$,\; $x=x$\;,\; $y=y$\; and\; $\eta_{s} = \frac{1}{2} \ln \left(\frac{t+z}{t-z}\right) $. $\eta_{s}$ is the space-time rapidity. These relations can easily be inverted to give  
\begin{equation}
    t\left(\tau, \eta_s\right) = \tau \cosh{\eta_s}\;, \quad  z\left(\tau, \eta_s\right) = \tau \sinh{\eta_s}.
\end{equation}\par
And the conversion of partial derivatives are given by,
\begin{align}
    \frac{\partial}{\partial t} &= \cosh{\eta_s}\frac{\partial}{\partial \tau} - \frac{\sinh{\eta_s}}{\tau}\frac{\partial}{\partial \eta_s},\\
    \frac{\partial}{\partial z}& = -\sinh{\eta_s}\frac{\partial}{\partial \tau} +\frac{\cosh{\eta_s}}{\tau}\frac{\partial}{\partial \eta_s}.
\end{align}
The terms containing $\sinh\eta_s$ as coefficient in above partial derivatives can be equated to zero, as at the end we will take mid-rapidity approximation which is $\eta_s \to 0$. 
Now, the four-velocity $u^\alpha = \gamma\left(1,0,0,z/\tau\right)= \gamma\left(1,0,0,\sinh\eta_s\right)$. Then the Lorentz factor $\gamma = \frac{1}{\sqrt{1-\sinh^2{\eta_s}}}$. With all these information in hand, we evaluate,
\begin{align*}
    \partial_\alpha u^\alpha &= \frac{\partial}{\partial t}\gamma + \frac{\partial}{\partial z}\frac{\gamma z}{\tau} \\
   & =  \cosh{\eta_s}\frac{\partial \gamma}{\partial \tau}+\frac{\cosh{\eta_s}}{\tau}\frac{\partial}{\partial \eta_s}\left(\gamma \sinh\eta_s\right).
\end{align*}
Then in the limit $\eta_s\to 0\;$,
\begin{equation}
    \partial_\alpha u^\alpha = \frac{1}{\tau}.
\end{equation}
\section{Calculation of $u^\alpha \partial_\alpha$}\label{AC}
\begin{align*}
    u^\alpha \partial_\alpha &= \gamma \frac{\partial}{\partial t} + \frac{\gamma z}{\tau} \frac{\partial}{\partial z}\\
    &= \gamma \cosh{\eta_s} \frac{\partial}{\partial\tau} + \frac{\gamma \sinh{\eta_s} \cosh{\eta_s}}{\tau} \frac{\partial}{\partial \eta_s} 
\end{align*}
Again in the mid-rapidity region,
\begin{equation}
    u^\alpha \partial_\alpha = \frac{\partial}{\partial\tau}.
\end{equation}

\section{Calculation of $u_\beta \xi^{\alpha \beta}=0$}\label{AD}
\begin{equation}
    u_\beta \xi^{\alpha \beta} = 2\eta u_\beta \lambda^{\alpha \beta}.
\end{equation}
Thus it is sufficient to show that $u_\beta \lambda^{\alpha \beta}$ vanishes.
\begin{equation}
  u_\beta \lambda^{\alpha \beta} = \frac{1}{2}\left(u_\beta \nabla^\alpha u^\beta + u_\beta \nabla^\beta u^\alpha \right)-\frac{1}{3} u_\beta \Delta^{\alpha \beta}\left(\nabla\cdot u\right) .
\end{equation}
The last term vanishes
which is a direct result from the definition of the projection operator.\par
Now the first term ,
\begin{equation}
    u_\beta\left(\nabla^\alpha u^\beta\right) = \frac{1}{2}\nabla^\alpha\left(u_\beta u^\beta\right) = 0.
\end{equation}
The second term,
\begin{equation}
    u_\beta\nabla^\beta u^\alpha = u^\beta \nabla_\beta u^\alpha \equiv a^\alpha\;,
\end{equation}
where $a^\mu$ is the four-acceleration of the fluid cell. But we are considering Bjorken flow, which corresponds to geodesic motion. Hence the fluid cells follow straight worldlines in space-time, thus they don't have any proper acceleration. Hence the second term also vanishes. Summing contributions from all three terms we get,
\begin{equation}
    u_\beta \xi^{\alpha \beta}=0.
\end{equation}
\section{Calculation of $u_\beta \partial_\alpha\xi^{\alpha\beta}$}\label{AE}
\begin{equation}
    u_\beta \partial_\alpha\xi^{\alpha\beta} = -2 \eta\lambda^{\alpha\beta} \partial_\alpha u_\beta.
\end{equation}
In Milne coordinates, i.e. $x^\alpha = \left(\tau, x,y, \eta_s\right)$, the metric tensor becomes,
\begin{equation}
    g^{\alpha \beta} = \text{diag}\left(1,-1, -1, -\tau^{-2}\right)\;, \quad g_{\alpha \beta} = \text{diag}\left(1,-1, -1, -\tau^{2}\right).
\end{equation}
And nonzero Christoffel symbols,
\begin{equation}
    \Gamma^\tau _{\eta_s \eta_s} = \tau \;, \quad     \Gamma^{\eta_s}   _{\tau \eta_s} =  \Gamma^{\eta_s} _{\eta_s \tau} = \frac{1}{\tau}.
\end{equation}
Now considering Bjorken flow in Milne coordinate, i.e. the velocity profile $u^\alpha = u_\alpha = \left(1,0,0,0\right)$, we calculate
\begin{itemize}
    \item Velocity gradients:
    \begin{equation}
        \nabla_\alpha u_\beta = \partial_\alpha u_\beta - \Gamma^\rho _{\alpha\beta}u_\rho = -\Gamma^\tau _{\alpha\beta}.
    \end{equation}
    Hence the only nonzero component is 
    \begin{equation}
        \nabla_{\eta_s} u_{\eta_s} = -\tau.
    \end{equation}
    Now raising indices we get,
    \begin{equation}\label{one}
        \nabla^{\eta_s} u^{\eta_s} = g^{\eta_s \eta_s} g^{\eta_s \eta_s }\nabla_{\eta_s} u_{\eta_s} = \left(-\tau^{-2}\right) \left(-\tau^{-2}\right) \left(-\tau\right) = -\frac{1}{\tau^3}.
    \end{equation}
    All other components of  $\nabla^\alpha u^\beta$ vanish. 
    \item Expansion scalar:
    \begin{equation}
        \nabla_ \cdot u = \nabla_\alpha u^\alpha = \partial_\alpha u^\alpha + \Gamma^\alpha _{\alpha \rho}u^\rho.
    \end{equation}
    The only nonzero term that survives is $\nabla_{\eta_s} u^{\eta_s} = \Gamma^{\eta_s} _{\eta_s \tau} u^\tau =  \frac{1}{\tau}$. So,
    \begin{equation}\label{two}
        \nabla \cdot u = \frac{1}{\tau}.
    \end{equation}
\end{itemize}
Now, realizing the projection tensor as $\Delta^{\alpha \beta} = \text{diag}\left(0,-1,-1,-\tau^{-2}\right)$, and using Eqs.~(\ref{one}, \ref{two}) we write,
\begin{equation}
    -2\eta \lambda^{\alpha \beta} \partial_\alpha u_\beta = -2\eta\lambda^{\eta_s \eta_s}\partial_\alpha u_\beta = \frac{4\eta}{3\tau^2} .
\end{equation}

\bibliography{apssamp_V1}

@PREAMBLE{
 "\providecommand{\noopsort}[1]{}" 
 # "\providecommand{\singleletter}[1]{#1}%" 
}

@article{Bazavov:2011nk,
    author = "Bazavov, A. and others",
    doi = "10.1103/PhysRevD.85.054503",
    journal = "Phys. Rev. D",
    volume = "85",
    pages = "054503",
    year = "2012"
}

@article{Shuryak:2014zxa,
    author = "Shuryak, Edward",
    doi = "10.1103/RevModPhys.89.035001",
    journal = "Rev. Mod. Phys.",
    volume = "89",
    pages = "035001",
    year = "2017"
}

@article{Rafelski:1982pu,
    author = "Rafelski, Johann and Muller, Berndt",
    doi = "10.1103/PhysRevLett.48.1066",
    journal = "Phys. Rev. Lett.",
    volume = "48",
    pages = "1066",
    year = "1982",
    note = "[Erratum: Phys.Rev.Lett. 56, 2334 (1986)]"
}

@article{Koch:1986ud,
    author = "Koch, P. and Muller, Berndt and Rafelski, Johann",
    doi = "10.1016/0370-1573(86)90096-7",
    journal = "Phys. Rept.",
    volume = "142",
    pages = "167--262",
    year = "1986"
}

@article{Koch:1988nn,
    author = "Koch, P. and Muller, Berndt and Stoecker, Horst and Greiner, W.",
    doi = "10.1142/S021773238800088X",
    journal = "Mod. Phys. Lett. A",
    volume = "3",
    pages = "737--742",
    year = "1988"
}

@article{Shuryak:1978ij,
    author = "Shuryak, Edward V.",
    reportNumber = "IYF-78-24",
    doi = "10.1016/0370-2693(78)90370-2",
    journal = "Phys. Lett. B",
    volume = "78",
    pages = "150",
    year = "1978"
}

@article{PHENIX:2004vcz,
    author = "Adcox, K. and others",
    collaboration = "PHENIX",
    doi = "10.1016/j.nuclphysa.2005.03.086",
    journal = "Nucl. Phys. A",
    volume = "757",
    pages = "184--283",
    year = "2005"
}

@article{Bellwied:2005kq,
    author = "Bellwied, R.",
    editor = "Jonson, B. and Meister, M. and Nyman, G. and Zhukov, M.",
    collaboration = "STAR",
    doi = "10.1016/j.nuclphysa.2005.02.141",
    journal = "Nucl. Phys. A",
    volume = "752",
    pages = "398--406",
    year = "2005"
}

@article{Muller:2006ee,
    author = "Muller, Berndt and Nagle, James L.",
    doi = "10.1146/annurev.nucl.56.080805.140556",
    journal = "Ann. Rev. Nucl. Part. Sci.",
    volume = "56",
    pages = "93--135",
    year = "2006"
}

@article{Gyulassy:2004zy,
    author = "Gyulassy, Miklos and McLerran, Larry",
    editor = "Rischke, D. and Levin, G.",
    doi = "10.1016/j.nuclphysa.2004.10.034",
    journal = "Nucl. Phys. A",
    volume = "750",
    pages = "30--63",
    year = "2005"
}

@article{Prasad:2022zbr,
    author = "Prasad, Suraj and Mallick, Neelkamal and Tripathy, Sushanta and Sahoo, Raghunath",
    doi = "10.1103/PhysRevD.107.074011",
    journal = "Phys. Rev. D",
    volume = "107",
    number = "7",
    pages = "074011",
    year = "2023"
}

@article{Prasad:2021bdq,
    author = "Prasad, Suraj and Mallick, Neelkamal and Behera, Debadatta and Sahoo, Raghunath and Tripathy, Sushanta",
    doi = "10.1038/s41598-022-07547-z",
    journal = "Sci. Rep.",
    volume = "12",
    number = "1",
    pages = "3917",
    year = "2022"
}

@article{Pal:2010es,
    author = "Pal, Subrata",
    doi = "10.1016/j.physletb.2010.01.017",
    journal = "Phys. Lett. B",
    volume = "684",
    pages = "211--215",
    year = "2010"
}

@article{Tannenbaum:2006ch,
    author = "Tannenbaum, M. J.",
    doi = "10.1088/0034-4885/69/7/R01",
    journal = "Rept. Prog. Phys.",
    volume = "69",
    pages = "2005--2060",
    year = "2006"
}

@article{Song:2007fn,
    author = "Song, Huichao and Heinz, Ulrich W.",
    reportNumber = "CERN-PH-TH-2007-154",
    doi = "10.1016/j.physletb.2007.11.019",
    journal = "Phys. Lett. B",
    volume = "658",
    pages = "279--283",
    year = "2008"
}

@article{ALICE:2010suc,
    author = "Aamodt, K and others",
    collaboration = "ALICE",
    reportNumber = "CERN-PH-EP-2010-059",
    doi = "10.1103/PhysRevLett.105.252302",
    journal = "Phys. Rev. Lett.",
    volume = "105",
    pages = "252302",
    year = "2010"
}

@article{ATLAS:2011ah,
    author = "Aad, Georges and others",
    collaboration = "ATLAS",
    reportNumber = "CERN-PH-EP-2011-124",
    doi = "10.1016/j.physletb.2011.12.056",
    journal = "Phys. Lett. B",
    volume = "707",
    pages = "330--348",
    year = "2012"
}

@article{Kharzeev:2014pha,
    author = "Kharzeev, Dmitri E.",
    doi = "10.1103/PhysRevD.90.074007",
    journal = "Phys. Rev. D",
    volume = "90",
    number = "7",
    pages = "074007",
    year = "2014"
}

@article{ALICE:2016fzo,
    author = "Adam, Jaroslav and others",
    collaboration = "ALICE",
    reportNumber = "CERN-EP-2016-153",
    doi = "10.1038/nphys4111",
    journal = "Nature Phys.",
    volume = "13",
    pages = "535--539",
    year = "2017"
}

@article{CMS:2016fnw,
    author = "Khachatryan, Vardan and others",
    collaboration = "CMS",
    reportNumber = "CMS-HIN-16-010, CERN-EP-2016-147",
    doi = "10.1016/j.physletb.2016.12.009",
    journal = "Phys. Lett. B",
    volume = "765",
    pages = "193--220",
    year = "2017"
}

@ARTICLE{Bjorken:2013boa,
   author       = "J. D. Bjorken, S. J. Brodsky and A. S. Goldhaber", 
   year         = "2013", 
   doi = "10.1016/j.physletb.2013.08.066",
   journal      = "Phys. Lett. B", 
   volume       = "726", 
   pages        = "344-346",
}

@article{PHENIX:2018lia,
    author = "Aidala, C. and others",
    collaboration = "PHENIX",
    doi = "10.1038/s41567-018-0360-0",
    journal = "Nature Phys.",
    volume = "15",
    number = "3",
    pages = "214--220",
    year = "2019"
}

@ARTICLE{Bozek:2012fw,
   author       = "P. Bozek and W. Broniowski", 
   year         = "2012", 
   doi = "10.1103/PhysRevC.85.044910",
   journal      = "Phys. Rev. C", 
   volume       = "85", 
   pages        = "044910",
}

@ARTICLE{Samanta:2023amp,
   author       = "R. Samanta and S. Bhatta and J. Jia and M. Luzum and J. Y. Ollitrault", 
   year         = "2024", 
   doi = "10.1103/PhysRevC.109.L051902",
   journal      = "Phys. Rev. C", 
   volume       = "109", 
   pages        = "L051902",
}

@ARTICLE{ALICE:2011ab,
   author       = "K. Aamodt \textit{et al.}", 
   collaboration = "ALICE",
   year         = "2011", 
    doi = "10.1103/PhysRevLett.107.032301",
   journal      = "Phys. Rev. Lett.", 
   volume       = "107", 
   pages        = "032301",
}

@ARTICLE{PHENIX:2011yyh,
   author       = "A. Adare \textit{et al.}", 
   collaboration = "PHENIX",
   year         = "2011", 
   doi = "10.1103/PhysRevLett.107.252301",
   journal      = "Phys. Rev. Lett.", 
   volume       = "107", 
   pages        = "252301",
}

@ARTICLE{CMS:2013wjq,
   author       = "S. Chatrchyan \textit{et al.}", 
   collaboration = "CMS",
   year         = "2014",  
   doi = "10.1103/PhysRevC.89.044906",
   journal      = "Phys. Rev. C", 
   volume       = "89", 
   pages        = "044906",
}

@ARTICLE{ATLAS:2019peb,
   author       = "M. Aaboud \textit{et al.}", 
   collaboration = "ATLAS",
   year         = "2020", 
doi = "10.1007/JHEP01(2020)051",
   journal      = "JHEP", 
   volume       = "01", 
   pages        = "051",
}

@ARTICLE{Heinz:2013th,
   author       = "U. Heinz and R. Snellings", 
   year         = "2013", 
   doi = "10.1146/annurev-nucl-102212-170540",
   journal      = "Ann. Rev. Nucl. Part. Sci.", 
   volume       = "63", 
   pages        = "123-151",
}

@ARTICLE{Broniowski:2009fm,
   author       = "W. Broniowski and M. Chojnacki and L. Obara", 
   year         = "2009", 
   doi = "10.1103/PhysRevC.80.051902",
   journal      = "Phys. Rev. C", 
   volume       = "80", 
   pages        = "051902",
}

@ARTICLE{Ollitrault:1992bk,
   author       = "J. Y. Ollitrault", 
   year         = "1992", 
   doi = "10.1103/PhysRevD.46.229",
   journal      = "Phys. Rev. D", 
   volume       = "46", 
   pages        = "229-245",
}

@ARTICLE{Yadav:2025vtc,
   author       = "A. K. Yadav and P. P. Bhaduri and S. Chattopadhyay", 
   year         = "2025",
   doi = "10.1140/epjc/s10052-025-14018-y",
   journal      = "Eur. Phys. J. C", 
   volume       = "85", 
   pages        = "247",
}

@ARTICLE{Shen:2011eg,
   author       = "C. Shen and U. Heinz and P. Huovinen and H. Song", 
   year         = "2011", 
  doi = "10.1103/PhysRevC.84.044903",
   journal      = "Phys. Rev. C", 
   volume       = "84", 
   pages        = "044903",
}

@ARTICLE{Hirano:2008hy,
   author       = "T. Hirano and N. van der Kolk and A. Bilandzic", 
   year         = "2010", 
   doi = "10.1007/978-3-642-02286-9_4",
   journal      = "Lect. Notes Phys.", 
   volume       = "785", 
   pages        = "139-178",
}

@ARTICLE{Huovinen:2006jp,
   author       = "P. Huovinen and P. V. Ruuskanen", 
   year         = "2006", 
   doi = "10.1146/annurev.nucl.54.070103.181236",
   journal      = "Ann. Rev. Nucl. Part. Sci.", 
   volume       = "56", 
   pages        = "163-206 ",
}

@ARTICLE{Niemi:2011ix,
   author       = "H. Niemi and G. S. Denicol and P. Huovinen and E. Molnar and D. H. Rischke", 
   year         = "2011", 
   doi = "10.1103/PhysRevLett.106.212302",
   journal      = "Phys. Rev. Lett.", 
   volume       = "106", 
   pages        = "212302",
}

@ARTICLE{Demir:2008tr,
   author       = "N. Demir and S. A. Bass", 
   year         = "2009", 
   doi = "10.1103/PhysRevLett.102.172302",
   journal      = "Phys. Rev. Lett.", 
   volume       = "102", 
   pages        = "172302",
}

@ARTICLE{Daher:2024vxk,
   author       = "A. Daher and L. Tinti and A. Jaiswal and R. Ryblewski", 
   year         = "2025", 
   doi = "10.1103/PhysRevD.111.074011",
   journal      = "Phys. Rev. D", 
   volume       = "111", 
   pages        = "074011",
}

@ARTICLE{Capellino:2022nvf,
   author       = "F. Capellino and A. Beraudo and A. Dubla and S. Floerchinger and S. Masciocchi and J. Pawlowski and I. Selyuzhenkov", 
   year         = "2022", 
   doi = "10.1103/PhysRevD.106.034021",
   journal      = "Phys. Rev. D", 
   volume       = "106", 
   pages        = "034021",
}

@ARTICLE{Yang:2023apw,
   author       = "Z. Yang and Y. Sun and L. W. Chen", 
   year         = "2024",   
   doi = "10.1103/PhysRevC.109.054907",
   journal      = "Phys. Rev. C", 
   volume       = "109", 
   pages        = "054907",
}

@ARTICLE{Meyer:2007ic,
   author       = "H. B. Meyer", 
   year         = "2007", 
   doi = "10.1103/PhysRevD.76.101701",
   journal      = "Phys. Rev. D", 
   volume       = "76", 
   pages        = "101701",
}

@ARTICLE{Grefa:2022sav,
   author       = "J. Grefa and M. Hippert and J. Noronha and J. Noronha-Hostler and I. Portillo and C. Ratti and R. Rougemont", 
   year         = "2022", 
   doi = "10.1103/PhysRevD.106.034024",
   journal      = "Phys. Rev. D", 
   volume       = "106", 
   pages        = "034024",
}

@ARTICLE{Teaney:2003kp,
   author       = "D. Teaney", 
   year         = "2003", 
   doi = "10.1103/PhysRevC.68.034913",
   journal      = "Phys. Rev. C", 
   volume       = "68", 
   pages        = "034913",
}

@ARTICLE{Ryu:2015vwa,
   author       = "S. Ryu and J. F. Paquet and C. Shen and G. S. Denicol and B. Schenke and S. Jeon and C. Gale", 
   year         = "2015", 
   doi = "10.1103/PhysRevLett.115.132301",
   journal      = "Phys. Rev. Lett.", 
   volume       = "115", 
   pages        = "132301",
}

@ARTICLE{Romatschke:2007mq,
   author       = "P. Romatschke and U. Romatschke", 
   year         = "2007", 
   doi = "10.1103/PhysRevLett.99.172301",
   journal      = "Phys. Rev. Lett.", 
   volume       = "99", 
   pages        = "172301",
}

@ARTICLE{Niemi:2015qia,
   author       = "H. Niemi and K. J. Eskola and R. Paatelainen", 
   year         = "2016", 
   doi = "10.1103/PhysRevC.93.024907",
   journal      = "Phys. Rev. C", 
   volume       = "93", 
   pages        = "024907",
}

@ARTICLE{Bernhard:2015hxa,
   author       = "J. E. Bernhard and P. W. Marcy and C. E. Coleman-Smith and S. Huzurbazar and R. L. Wolpert and S. A. Bass", 
   year         = "2015", 
   doi = "10.1103/PhysRevC.91.054910",
   journal      = "Phys. Rev. C", 
   volume       = "91", 
   pages        = "054910",
}

@ARTICLE{Heffernan:2023utr,
   author       = "M. R. Heffernan and C. Gale and S. Jeon and J. F. Paquet", 
   year         = "2024", 
   doi = "10.1103/PhysRevC.109.065207",
   journal      = "Phys. Rev. C", 
   volume       = "109", 
   pages        = "065207",
}

@ARTICLE{Nijs:2022rme,
   author       = "G. Nijs and W. van der Schee", 
   year         = "2022", 
   doi = "10.1103/PhysRevLett.129.232301",
   journal      = "Phys. Rev. Lett.", 
   volume       = "129", 
   pages        = "232301",
}

@ARTICLE{Parkkila:2021yha,
   author       = "J. E. Parkkila and A. Onnerstad and S. F. Taghavi and C. Mordasini and A. Bilandzic and M. Virta and D. J. Kim", 
   year         = "2022", 
   doi = "10.1016/j.physletb.2022.137485",
   journal      = "Phys. Lett. B", 
   volume       = "835", 
   pages        = "137485",
}

@ARTICLE{Chaudhuri:2009uk,
   author       = "A. K. Chaudhuri", 
   year         = "2009", 
   doi = "10.1016/j.physletb.2009.10.068",
   journal      = "Phys. Lett. B", 
   volume       = "681", 
   pages        = "418-422",
}

@ARTICLE{Chaudhuri:2010in,
   author       = "A. K. Chaudhuri", 
   year         = "2010", 
   doi = "10.1103/PhysRevC.82.047901",
   journal      = "Phys. Rev. C", 
   volume       = "82", 
   pages        = "047901",
}

@ARTICLE{Gubser:2012gy,
   author       = "S. S. Gubser", 
   year         = "2013",
   doi = "10.1103/PhysRevC.87.014909",
   journal      = "Phys. Rev. C", 
   volume       = "87", 
   pages        = "014909",
}

@ARTICLE{Bagchi:2023ysc,
   author       = "A. Bagchi and K. S. Kolekar and A. Shukla", 
   year         = "2023,", 
   doi = "10.1103/PhysRevLett.130.241601",
   journal      = "Phys. Rev. Lett.", 
   volume       = "130", 
   pages        = "241601",
}

@ARTICLE{Ciambelli:2018xat,
   author       = "L. Ciambelli and C. Marteau and A. C. Petkou and P. M. Petropoulos", 
   year         = "2018", 
   doi = "10.1088/1361-6382/aacf1a",
   journal      = "Class. Quant. Grav.", 
   volume       = "35", 
   pages        = "165001",
}

@ARTICLE{Petkou:2022bmz,
   author       = "A. C. Petkou and P. M. Petropoulos and D. R. Betancour and K. Siampos", 
   year         = "2022",  
   doi = "10.1007/JHEP09(2022)162",
   journal      = "JHEP", 
   volume       = "09", 
   pages        = "162",
}

@ARTICLE{Bjorken:1982qr,
   author       = "J. D. Bjorken", 
   year         = "1983",  
   doi = "10.1103/PhysRevD.27.140",
   journal      = "Phys. Rev. D", 
   volume       = "27", 
   pages        = "140-151",
}

@ARTICLE{ALICE:2018cpu,
   author       = "S. Acharya \textit{et al.}", 
   collaboration = "ALICE",
   year         = "2019", 
   doi = "10.1016/j.physletb.2018.12.048",
   journal      = "Phys. Lett. B", 
   volume       = "790", 
   pages        = "35-48",
}

@book{Weinberg:1972kfs,
    author = "Weinberg, Steven",
    title = "{Gravitation and Cosmology}: {Principles and Applications of the General Theory of Relativity}",
    isbn = "978-0-471-92567-5, 978-0-471-92567-5",
    publisher = "John Wiley and Sons",
    address = "New York",
    year = "1972"
}

@article{Dutta:1999cn,
    author = "Dutta, D. and Mohanty, A. K. and Kumar, K. and Choudhury, R. K.",
    doi = "10.1103/PhysRevC.61.034902",
    journal = "Phys. Rev. C",
    volume = "61",
    pages = "034902",
    year = "2000"
}

@article{Laermann:2003cv,
    author = "Laermann, Edwin and Philipsen, Owe",
    doi = "10.1146/annurev.nucl.53.041002.110609",
    journal = "Ann. Rev. Nucl. Part. Sci.",
    volume = "53",
    pages = "163--198",
    year = "2003"
}

@article{PHOBOS:2004zne,
    author = "Back, B. B. and others",
    collaboration = "PHOBOS",
     doi = "10.1016/j.nuclphysa.2005.03.084",
    journal = "Nucl. Phys. A",
    volume = "757",
    pages = "28--101",
    year = "2005"
}

@CONTROL{REVTEX42Control}

@CONTROL{apsrev42Control,author="08",editor="1",pages="0",title="0",year="1"}

\end{document}